\documentclass[floatfix,twocolumn,linenumbers]{aastex63}
\begin{document}
\nolinenumbers


\shorttitle{
Cosmological-Scale Ly$\alpha$ forest absorption Around Galaxies and AGN
}

\shortauthors{
Sun et al.
}

\title{
Cosmological-Scale Lyman-alpha Forest Absorption Around Galaxies and AGN\\ 
Probed with the HETDEX and SDSS Spectroscopic Data
}

\correspondingauthor{Dongsheng Sun}
\email{sunds@icrr.u-tokyo.ac.jp}

\author[0000-0002-1199-6523]{Dongsheng Sun}
\affiliation{Institute for Cosmic Ray Research, The University of Tokyo, 5-1-5 Kashiwanoha, Kashiwa, Chiba 277-8582, Japan}
\affiliation{Department of Astronomy, Graduate School of Science, the University of Tokyo, 7-3-1 Hongo, Bunkyo, Tokyo 113-0033, Japan}

\author[0000-0003-4985-0201]{Ken Mawatari}
\affiliation{National Astronomical Observatory of Japan, 2-21-1 Osawa, Mitaka, Tokyo 181-8588, Japan}
\affiliation{Institute for Cosmic Ray Research, The University of Tokyo, 5-1-5 Kashiwanoha, Kashiwa, Chiba 277-8582, Japan}

\author[0000-0002-1049-6658]{Masami Ouchi}
\affiliation{National Astronomical Observatory of Japan, 2-21-1 Osawa, Mitaka, Tokyo 181-8588, Japan}
\affiliation{Institute for Cosmic Ray Research, The University of Tokyo, 5-1-5 Kashiwanoha, Kashiwa, Chiba 277-8582, Japan}
\affiliation{Kavli Institute for the Physics and Mathematics of the Universe (Kavli IPMU, WPI), The University of Tokyo, 5-1-5 Kashiwanoha, Kashiwa, Chiba, 277-8583, Japan}

\author[0000-0001-9011-7605]{Yoshiaki Ono}
\affiliation{Institute for Cosmic Ray Research, The University of Tokyo, 5-1-5 Kashiwanoha, Kashiwa, Chiba 277-8582, Japan}

\author[0000-0002-1319-3433]{Hidenobu Yajima}
\affiliation{Center for Computational Sciences, University of Tsukuba, Ten-nodai, 1-1-1 Tsukuba, Ibaraki 305-8577, Japan}

\author[0000-0003-3817-8739]{Yechi Zhang}
\affiliation{Institute for Cosmic Ray Research, The University of Tokyo, 5-1-5 Kashiwanoha, Kashiwa, Chiba 277-8582, Japan}
\affiliation{Department of Astronomy, Graduate School of Science, the University of Tokyo, 7-3-1 Hongo, Bunkyo, Tokyo 113-0033, Japan}
\affiliation{Kavli Institute for the Physics and Mathematics of the Universe (Kavli IPMU, WPI), The University of Tokyo, 5-1-5 Kashiwanoha, Kashiwa, Chiba, 277-8583, Japan}

\author{Makito Abe}
\affiliation{Center for Computational Sciences, University of Tsukuba, Ten-nodai, 1-1-1 Tsukuba, Ibaraki 305-8577, Japan}

\author[0000-0003-4381-5245]{William P. Bowman}
\affiliation{Department of Astronomy, Yale University, New Haven, CT 06520}

\author[0000-0002-2307-0146]{Erin Mentuch Cooper}
\affiliation{Department of Astronomy, The University of Texas at Austin, 2515 Speedway Boulevard, Austin, TX 78712, USA}
\affiliation{McDonald Observatory, The University of Texas at Austin, 2515 Speedway Boulevard, Austin, TX 78712, USA}

\author[0000-0002-8925-9769]{Dustin Davis}
\affiliation{Department of Astronomy, The University of Texas at Austin,
2515 Speedway Boulevard, Austin, TX 78712, USA}

\author[0000-0003-2575-0652]{Daniel J. Farrow}
\affiliation{University Observatory, Fakult\"at f\"ur Physik, Ludwig-Maximilians University Munich, Scheinerstrasse 1, 81679 Munich, Germany}
\affiliation{Max-Planck Institut f\"ur extraterrestrische Physik, Giessenbachstrasse 1, 85748 Garching, Germany}

\author[0000-0002-8433-8185]{Karl Gebhardt}
\affiliation{Department of Astronomy, The University of Texas at Austin, 2515 Speedway Boulevard, Austin, TX 78712, USA}

\author[0000-0001-6717-7685]{Gary J. Hill}
\affiliation{McDonald Observatory, The University of Texas at Austin, 2515 Speedway Boulevard, Austin, TX 78712, USA}
\affiliation{Department of Astronomy, The University of Texas at Austin, 2515 Speedway Boulevard, Austin, TX 78712, USA}

\author[0000-0001-5561-2010]{Chenxu Liu}
\affiliation{South-Western Institute for Astronomy Research, Yunnan University, Kunming, Yunnan, 650500, People’s Republic of China}
\affiliation{Department of Astronomy, The University of Texas at Austin, 2515 Speedway Boulevard, Austin, TX 78712, USA}

\author[0000-0001-7240-7449]{Donald P. Schneider}
\affiliation{Department of Astronomy \& Astrophysics, The Pennsylvania State University, University Park, PA 16802, USA}
\affiliation{Institute for Gravitation and the Cosmos, The Pennsylvania State University, University Park, PA 16802, USA}


\begin{abstract}
\nolinenumbers
We present cosmological-scale 3-dimensional (3D) neutral hydrogen ({\sc Hi}) tomographic maps at $z=2-3$ over a total of 837 deg$^2$ in two blank fields 
that are developed with Ly$\alpha$ forest absorptions of 14,736 background Sloan Digital Sky Survey (SDSS) quasars at $z$=2.08-3.67.
Using the tomographic maps, we investigate 
the large-scale ($\gtrsim 10$ $h^{-1}$cMpc) 
average {\sc Hi} radial profiles and two-direction profiles of the line-of-sight (LoS) and transverse (Trans) directions around galaxies and AGN at $z=2-3$ identified by the Hobby-Eberly Telescope Dark Energy eXperiment (HETDEX) and SDSS surveys, respectively.
The peak of the {\sc Hi} radial profile around galaxies is lower than the one around AGN, suggesting that the dark-matter halos of galaxies are less massive on average than those of AGN.
%
The LoS profile of AGN is narrower than the Trans profile, indicating the Kaiser effect. There exist weak absorption outskirts at $\gtrsim 30$ $h^{-1}$cMpc beyond {\sc Hi} structures of galaxies and AGN found in the LoS profiles that can be explained by the {\sc Hi} gas at $\gtrsim 30$ $h^{-1}$cMpc falls toward the source positions.
Our findings indicate that the {\sc Hi} radial profile of AGN has transitions from proximity zones ($\lesssim$ a few $h^{-1}$cMpc) to the {\sc Hi} structures ($\sim 1-30$ $h^{-1}$cMpc) and the weak absorption outskirts ($\gtrsim 30$ $h^{-1}$cMpc).
Although there is no significant dependence of AGN types (type-1 vs. type-2) on the {\sc Hi} profiles, the peaks of the radial profiles anti-correlate with AGN luminosities, suggesting that AGN's ionization effects are stronger than the gas mass differences.
\end{abstract}

\keywords{galaxies: formation --- galaxies: evolution --- galaxies: high-redshift --- galaxies: halos --- intergalactic medium}


\section{Introduction} \label{sec:intro}
Galaxy formation in the Universe is closely related to the neutral hydrogen ({\sc Hi}) gas in the intergalactic medium (IGM).
Within the modern paradigm of galaxy formation, galaxies form and evolve in the filament structure of {\sc Hi} gas (e.g., \citealt{meiksin+09}; \citealt{mo+10}). 
 Cosmological hydrodynamics simulations suggest that the picture of galaxy formation and evolution is associated with large-scale baryonic gas exchange between the galaxy and the IGM (\citealt{fox+17}; \citealt{van+17}).

The circulation of gas is one of the keys to understanding galaxy formation and evolution.
The interplay of gravitational and feedback-driven processes can have surprisingly large effects on the large scale behavior of the IGM.
Some of the radiation produced by massive stars and black hole accretion disks can escape from the dense gaseous environments and propagate out of galaxies and photoionize the {\sc Hi} gas in the circumgalactic medium (CGM) and even in the IGM \citep{mukae+20,astro2021}.

Great progress has been achieved in exploring the Ly$\alpha$ forest absorption around galaxies and active galactic nuclei (AGN).
The cross-correlation of the {\sc Hi} in the IGM and galaxies has been detected by Ly$\alpha$ absorption features in the spectra of background quasars (e.g., \citealt{rauch+98}; \citealt{Faucher-Giguere+08}; \citealt{Prochaska+13}) and bright star-forming galaxies 
\citep{Steidel+10,Mawatari+16,Thomas+17}.
The Keck Baryon Structure Survey \citep[KBSS:][]{Rudie+12,Rakic+12,Turner+14}, the Very Large Telescope LBG Redshift Survey \citep[VLRS:][]{Crighton+11,Tummuangpak+14}, and other spectroscopic programs \citep[e.g.,][]{Adelberger+03,Adelberger+05} have investigated the detailed properties of the Ly$\alpha$ forest absorption around galaxies.
These observations target {\sc Hi} gas around galaxies on the scale of the circumgalactic medium (CGM).
Recently, 3-dimensional (3D) {\sc Hi} tomography mapping, a powerful technique to reconstruct the large scale structure of {\sc Hi} gas, has been developed by \cite{Lee+14a,Lee+16,Lee+18}.
{\sc Hi} tomography mapping is originally proposed by \cite{Pichon+01} and \cite{Caucci+08} with the aim of reconstructing the 3D matter distribution from the Ly$\alpha$ transmission fluctuation of multiple sightlines. By this technique, the COSMOS Ly$\alpha$ Mapping and Tomography Observations (CLAMATO) survey \citep{Lee+14a,Lee+18} has revealed {\sc Hi} large scale structures with spatial resolutions of 2.5 $h^{-1}$ comoving Megaparsec (cMpc). This survey demonstrates the power of 3D {\sc Hi} tomography mapping in a number of applications, including the study of a protocluster at $z = 2.44$ \citep{Lee+16} and the identification of cosmic voids \citep{Krolewski+18}.
Due to an interpolation algorithm (Section \ref{subsec:reconstruction}) used in the reconstruction of the 3D {\sc Hi} tomography map, we are able to estimate the Ly$\alpha$ forest absorption along lines-of-sight where there are no available background sources.
Based on the 3D {\sc Hi} tomography map of the CLAMATO survey, \cite{momose+21} have reported measurements the IGM {\sc Hi}–galaxy cross-correlation function (CCF) for several galaxy populations.
Due to the limited volume of the CLAMATO 3D IGM tomography data, \cite{momose+21} cannot construct the CCFs at scales over 24 $h^{-1}$cMpc in the direction of transverse to the line-of-sight.
\cite{mukae+20} have investigated a larger field than the one of \cite{momose+21} using 3D {\sc Hi} tomography mapping and report that a huge ionized structure of {\sc Hi} gas associated with an extreme QSO overdensity region in the EGS field.
\cite{mukae+20} interpret the large ionized structure as the overlap of multiple proximity zones which are photoionized regions created by the enhanced ultraviolet background (UVB) of quasars.
However, \cite{mukae+20} found only one example of a huge ionized bubble, and no others have been reported in the literature.

Dispite the great effort made by previous studies, the limited volume of previous work prevents us from understanding how ubiquitous or rare these large ionized structures are. In order to answer this question, we must investigate the statistical Ly$\alpha$ forest absorptions around galaxies and AGN at much larger spatial scales ($\gtrsim10~h^{-1}$cMpc). Although \cite{momose+21} derived CCFs for different populations: Ly$\alpha$ emitters (LAEs), H$\alpha$ emitters (HAEs), [{\sc Oiii}] emitters (O3Es), active galactic nuclei (AGN), and submillimeter galaxies (SMGs), on a scale of more than $20$ $h^{-1}$cMpc, the limited sample size results in large uncertainties in the CCF at large scales and prevents definitive conclusions to be made regarding the statistical Ly$\alpha$ forest absorptions around galaxies and AGN.

Another open question is the luminosity and AGN type dependence of the large scale Ly$\alpha$ forest absorption around AGN.
\cite{FR+13} have estimated the Ly$\alpha$ forest absorption around AGN using the Sloan Digital Sky Survey (SDSS; \citealt{York+00}) data release 9 quasar catalog (DR9Q; \citep{paris+11}) and find no dependence of the Ly$\alpha$ forest absorption on AGN luminosity.
In this study, we investigate the luminosity dependence using the SDSS data release 14 quasar (DR14Q; \citealt{Paris+18}) catalog, which includes sources $\sim 2$ magnitude fainter than those used by \cite{FR+13}.
In the AGN unification model (\citealt{Antonucci+85}; see also \citealt{Spinoglio+21}), which provides a physical picture that a hot accretion disk of super-massive blackhole is obscured by a dusty torus, the type-1 and type-2 classes are produced by different accretion disk viewing angles.
In this picture, the type-1 (type-2) AGN is biased to AGN with a wide (narrow) opening angle.
In the case of type-1 AGN, one can directly observe the accretion disks and the broad line region, while for type-2 AGN, only the narrow line region is observable.
Previous studies have identified the proximity effect that the IGM of type-1 AGN is statistically more ionized due to the local enhancement of the UV background on the line-of-sight passing near the AGN \citep{FaucherGigu+08}.
Based on the unification model, the type-2 AGN obscured on the line of sight statistically radiates in transverse direction.
The investigation of the AGN type dependence on the surrounding {\sc Hi} can reveal the large scale Ly$\alpha$ forest absorption influenced by the direction of radiation from the AGN.


To investigate the Ly$\alpha$ forest absorptions around galaxies and AGN on large scales, over tens of $h^{-1}$cMpc, we need conduct a new study in a field with length of any side larger than 100 $h^{-1}$cMpc.
We reconstruct a 3D {\sc Hi} tomography maps of Ly$\alpha$ forest absorption at $z \sim$ $2-3$ in a total area of $837$ deg$^2$.
We use $\gtrsim 15,000$ background sightlines from SDSS quasars \citep{Paris+18,Lyke+20} for the {\sc Hi} tomography map reconstruction and have a large number of unbiased galaxies and AGN from the Hobby Eberly Telescope Dark Energy eXperiment (HETDEX; \citealt{Gebhardt+21}) and SDSS surveys for the investigations of the large scale Ly$\alpha$ forest absorptions around galaxies and AGN.

This paper is organized as follows. 
Section \ref{sec:data} describes the details of the HETDEX survey and our spectroscopic data.
Our foreground and background samples of galaxies and AGN are presented in Section \ref{sec:sample}.
The technique of creating the {\sc Hi} tomography mapping and the reconstructed {\sc Hi} tomography map
are described in Section \ref{sec:hi_tomography}, and the 
observational results of Ly$\alpha$ forest absorptions around galaxies and AGN are given in Section \ref{sec:result}.
In this section, we also interpret our results in the context of previous studies, and investigate the dependence of out tomography maps on AGN type and luminosity.
We adopt a cosmological parameter set of ($\Omega_m$, $\Omega_{\rm \Lambda}$, $h$) = (0.29, 0.71, 0.7) in this study.

\section{Data} \label{sec:data}
\subsection{HETDEX Spectra} \label{subsec:hetdex}
HETDEX provides an un-targeted, wide-area, integral field spectroscopic survey, and aims to determine the evolution of dark energy in the redshift range $1.88-3.52$ using $\sim1$ million Lyman-$\alpha$ emitters (LAEs) over 540 deg$^2$ in the northern and equatorial fields that are referred to as ``Spring'' and ``Fall'' fields, respectively. The total survey volume is $\sim 10.9$ comoving Gpc$^3$.

The HETDEX spectroscopic data are gathered using the 10 m Hobby-Eberly Telescope \citep[HET;][]{lwr94,Hill+21} to collect light for
the Visible Integral-field Replicable Unit Spectrograph \citep[VIRUS;][]{hil18a,Hill+21} with 78 integral field unit \citep[IFUs;][]{kelz14} fiber arrays. VIRUS covers a wavelength, with resolving power ranging from $750-950$. 
Each IFU has 448 fibers with a $1''.5$ diameter. The $78$ IFUs are spread over the $22$ arcmin field of view, with a $1/4.6$ fill factor.
Here we make use of the data release 2 of the HETDEX \citep[HDR2;][]{Cooper+23} over the Fall and Spring fields. In this study, we investigate the fields where HETDEX survey data are taken between 2017 January and 2020 June.
The effective area is 11542 arcmin$^2$. 
The estimated depth of an emission line at S/N$=5$ reaches $3-4 \times 10^{-17}$ erg cm$^{-2}$ s$^{-1}$.



\subsection{Subaru HSC Imaging} \label{subsec:hsc}
The HETDEX-HSC imaging survey was carried out in a total time allocation of 3 nights in $2015-2018$ (semesters S15A, S17A, and S18A; PI: A. Schulze) and $2019-2020$ (semester S19B; PI: S. Mukae) over a $\sim$250 deg$^2$ area in the Spring field, accomplishing a 5$\sigma$ limiting magnitude of $r = 25.1$ mag.
The SSP-HSC program has obtained deep multi-color imaging data on
the 300 deg$^2$ sky, half of which overlaps with the HETDEX footprints. In this study, we use the $r$-band imaging data from the public data release 2 (PDR2) of SSP-HSC. The 5$\sigma$ depth of the SSP-HSC PDR2 $r$-band imaging data is typically $27.7$ mag for the $3''.0$ diameter aperture. The data reduction of HETDEX-HSC survey and SSP-HSC program are processed with HSC pipeline software, \texttt{hscPipe} \citep{bosch18} version $6.7$.

Because the spectral coverage width of the HETDEX survey is narrow, only 2000 \AA, most sources appear as single-line emitters. 
Furthermore, since the {\sc Oii} doublet is not resolved, we rely on the equivalent width (EW) to distinguish Ly$\alpha$ from {\sc Oii}.
 The high-$z$ Ly$\alpha$ emission is typically stronger than low-$z$ {\sc [Oii]} lines, due to the intrinsic line strengths and the cosmological effects.
%
%
The continuum estimate from the HETDEX spectra reach about g$=25.5$ \citep{Davis+21,Cooper+23} and we improve on this using the deep HSC imaging. We estimate EW using continua measured from two sets of images taken by HSC r-band imaging survey for HETDEX (HETDEX-HSC survey) and the Subaru Strategic Program \citep[SSP-HSC;][]{aihara18b}.
\citeauthor{Davis+21} and \citeauthor{Cooper+23} find that our contamination of {\sc Oii} emitters in the LAE sample to be below 2\%.

\subsection{SDSS-IV eBOSS Spectra} \label{subsec:sdss}
We use quasar data from  eBOSS \citep{Dawson+16}, which is publically available in the SDSS Data Release 14 and 16 quasar catalog \citep[DR14Q, DR16Q;][]{Paris+18,Lyke+20}.
The cosmology survey, eBOSS, is part of SDSS-IV. The eBOSS quasar targets are selected by the XDQSOz
method \citep{Bovy+12} and the color cut
\begin{equation} \label{eq:quasar_select}
    m_{opt}-m_{WISE}\geq(g-i)+3,
\end{equation}
where $m_{opt}$ is a weighted stacked magnitude in the $g, r$ and $i$ bands and $m_{WISE}$ is a weighted stacked magnitude in the W1 and W2 bands of the Wide-Field Infrared Survey (WISE; \citealt{Wright+10}). 
The aim of the eBOSS is to accomplish precision angular-diameter distance measurements and the Hubble parameter determination at $z \sim 0.6-3.5$ using different tracers of the underlying density fields over 7500 deg$^2$. Its final goal is to obtain spectra of $\sim 2.5$ million luminous red galaxies, $\sim 1.95$ million emission line galaxies, $\sim$ 450,000 QSOs at
$0.9 \leq z \leq 2.2$, and the Lyman-$\alpha$ forest of 60,000 QSOs at $z>2$ over four years of operation.

The eBOSS program is conducted with twin SDSS spectrographs \citep{Smee13}, which are fed by 1,000 fibers connected from the focal plane of the 2.5m Sloan telescope \citep{Gunn06} at Apache Point Observatory. SDSS spectrographs have a fixed spectral bandpass of $3600-10000$ \AA \ over the 7 deg$^2$ field of view. The spectral resolution varies from 1300 at the blue end to 2600 at the red end,
where one pixel corresponds to $1.8-5.2$ \AA.



\section{Samples} \label{sec:sample}

Our study aims to map the statistical distribution of {\sc Hi} gas on a cosmological scale around foreground galaxies and AGN by the 3D {\sc Hi} tomography mapping technique with background sources at $z=2-3$. 
We use the foreground galaxies, foreground AGN, and background sources presented in Sections \ref{subsec:fg_galaxy}, \ref{subsec:fg_agn}, and \ref{subsec:bkagn}, respectively.

\begin{figure*}[ht!]
\begin{center}
\includegraphics[scale=0.55]{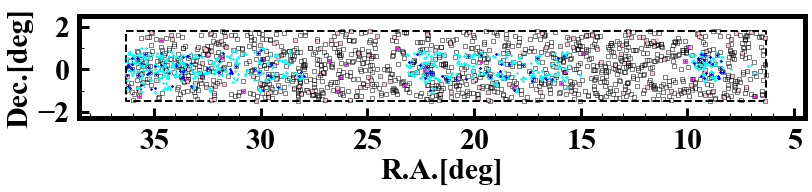}
\end{center}
\caption{
Sky distribution of the foreground AGN and galaxies at ${\rm z}=2.0-2.2$ in the ExFall field. The squares present the positions of All-AGN sample sources. Pink (magenta) squares represent the sources of the T1-AGN (T2-AGN) sample. The cyan and blue dots show the positions of the Galaxy and T1-AGN(H) sample sources, respectively. The black dashed line indicates the border of the {\sc Hi} tomography map in the Exfall field.}
\label{fig:our_study_field}
\end{figure*}

Two of the  goals of this study are to explore the dependence of luminosity and AGN type on the Ly$\alpha$ forest absorption. To examine statistical results, we need a large number of bright AGN and type-2 AGN. Compared to moderately bright AGN and type-1 AGN, bright AGN and type-2 AGN are relatively rare.
To obtain a sufficiently large samples of bright AGN and type-2 AGN, we expand the Spring and Fall fields of the HETDEX survey, from which we are able to investigate the statistical luminosity and AGN type dependence of the HI distribution around AGN (Section \ref{subsec:fg_agn}).
The northern extended Spring field flanking the HETDEX survey fields, referred to as the ``ExSpring field", covers over 738 deg$^2$, while the equatorial extended Fall field flanking the HETDEX survey fields, here after ``ExFall field", covers 99 deg$^2$.
The total area of our 3D {\sc Hi} tomography mapping field is 837 deg$^2$ in the ExSpring and ExFall fields that is referred to as ``our study field".
Our analysis is conducted in our study field where the foreground galaxies+AGN and the background sources overlap on the sky. As an example, we present the foreground galaxies+AGN in the ExFall field at ${\rm z}=2.0-2.2$ in Figure \ref{fig:our_study_field}. We also present the sky distribution of the background sources within the ExFall field in Figure \ref{fig:exfall_bk}. The rest of the foreground and background sources are shown in the Appendix.

\begin{figure*}[ht!]
\begin{center}
\includegraphics[scale=0.54]{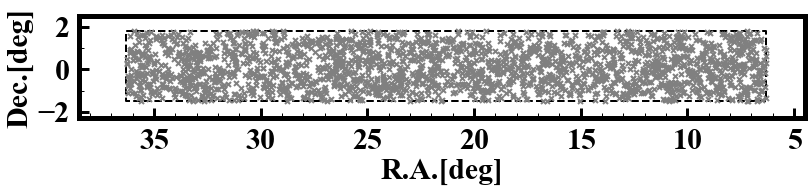}
\end{center}
\caption{Sky distribution of background AGN in the ExFall field. The gray crosses indicate background AGN that are used to reconstruct our {\sc Hi} tomography map. The back dashed line has the same meaning as that in Figure \ref{fig:our_study_field}.}
\label{fig:exfall_bk}
\end{figure*}


\begin{deluxetable*}{cccccccc}[ht!]
\tablecaption{Sample size of foreground samples at $z = 2-3$ \label{tab:fg}} 
\tablecolumns{4}
\tablewidth{0pt}
\tablehead{
\colhead{Name of sample} &
\colhead{ExFall} &
\colhead{ExSpring} &
\colhead{Total} &
\colhead{Survey} &
\colhead{Criteria} 
}
\startdata
Galaxy & 3431 & 11436
 & 14867 & HETDEX & EW$_0>20$ \AA, FWHM$_{\rm Ly\alpha}$ $<1000$ km/s, Muv$>-22$ mag \\ 
T1-AGN(H) & 438 & 1349 & 1787 & HETDEX & EW$_0>20$ \AA, FWHM$_{\rm Ly\alpha}$ $>1000$ km/s \\
T1-AGN & 2393 & 12300
 & 14693 & SDSS & FWHM$_{\rm Ly\alpha}$ $>1000$ km/s \\ 
T2-AGN & 436 & 1633
 & 2069 & SDSS & FWHM$_{\rm Ly\alpha}$ $<1000$ km/s \\ 
\enddata
\end{deluxetable*}

\begin{deluxetable*}{cccccccc}[ht!]
\tablecaption{Sample size of background sample at $z = 2.08-3.67$ \label{tab:bg}} 
\tablecolumns{4}
\tablewidth{0pt}
\tablehead{
\colhead{Name of sample} &
\colhead{ExFall} &
\colhead{ExSpring} &
\colhead{Total} &
\colhead{Survey} &
\colhead{Criteria} 
}
\startdata
background AGN & 2181 & 12555
 & 14736 & SDSS & $\langle S/N \rangle_{\rm Ly\alpha forest}>1.4$
\\ 
\enddata
\end{deluxetable*}

\subsection{Foreground Galaxy Sample} \label{subsec:fg_galaxy}
%
We make a sample of foreground galaxies from the data of the HETDEX spectra (Section \ref{subsec:hetdex}) and the Subaru HSC images (Section \ref{subsec:hsc}). 
With these data, \cite{Zhang21} have build a catalog of LAEs that have the rest-frame equivalent widths (${\rm EW}_0$) of ${\rm EW}_0>20$ \AA \ and the HETDEXs Emission Line eXplorer (ELiXer) probabilities \citep{Davis+21,Davis+23} larger than 1. This ${\rm EW}_0$ cut is similar to previous LAE studies (e.g., \citealt{gronwall07,konno16}).
This catalog of LAEs is composed of 15959 objects.
Because the LAE catalog of \cite{Zhang21} consists of galaxies, type-1 AGN, and type-2 AGN, we isolate galaxies from the sources of the LAE catalog with the limited observational quantities, Ly$\alpha$ and UV magnitude ($M_{\rm UV}$), that can be obtained from the HETDEX and Subaru/HSC data.
Because type-1 AGN have broad-line Ly$\alpha$ emission, we remove sources with broad-line Ly$\alpha$ whose full width half maximum (FWHM) of the Ly$\alpha$ emission lines are greater than 1000 km s$^{-1}$. To remove clear type-2 AGN from the LAE catalog, we apply a UV magnitude cut of $M_{\rm UV}>-22$ mag that is the bright end of the UV luminosity function dominated by star-forming galaxies \citep{Zhang21}.
We then select sources in our study field, and apply the redshift cut of $z=2.0-3.0$ (as measured by the principle component analysis of multiple lines; \citealt{Paris+18}) to match the redshift range over which we construct {\sc Hi} tomography map. These redshifts are measured with Ly$\alpha$ emission \citep{Zhang21}, because Ly$\alpha$ is the only emission available for all of the sources.

By these selections, we obtain 14130 star-forming galaxies from the LAE catalog. These 14130 star-forming galaxies are referred to as the ``galaxy" sample in this study.

\subsection{Foreground AGN Samples} \label{subsec:fg_agn}
In this subsection, we describe how we select foreground 
AGN from two sources, (a) the combination of the HETDEX spectra and the HSC imaging data and (b) the SDSS DR14Q catalog. The type-$1$ AGN are identified with the sources of (a) and (b), while the type-$2$ AGN are drawn from the source of (b).

With the source (a) that is the same as the one stated in Section \ref{subsec:fg_galaxy}, \cite{Zhang21} have constructed the LAE catalog.
We use the catalog of \cite{Zhang21} to select LAEs at $z\sim2-3$ that fall in our study field.
Applying a Ly$\alpha$ line width criterion of FWHM $>1000$ km s$^{-1}$ with the HETDEX spectra, we identify broad-line AGN, i.e. type-1 AGN, from the LAEs. 
We thus obtain 1829 type-1 AGN that are referred to as T1-AGN(H).

We use the width of Ly$\alpha$ emission line for the selection of type-1 AGN. This is because no other emission lines characterising AGN, e.g. {\sc Civ}, are available for all of the LAEs due to the limited wavelength coverage and the sensitivity of HETDEX. Similarly, the redshifts of T1-AGN(H) objects are measured with Ly$\alpha$ emission whose redshifts may be shifted from the systemic redshifts by up to a few 100 km s$^{-1}$ (See Section \ref{subsec:fg_galaxy}). We do not select type-2 AGN from the source of (a), because we cannot identify type-2 AGN easily with the given data set of source (a).

From the source (b), we obtain the other samples of foreground AGN. We first choose objects with a classification of QSOs of the SDSS DR14Q, and remove objects outside the redshift range of $z=2.0-3.0$ in 
our study field. 
We obtain 23721 AGN.
For 16762 out of 23721 AGN, Ly$\alpha$ FWHM measurements are available from \cite{Rakshit20}.
The other AGN without FWHM measurement are removed due to the poor quality of the Ly$\alpha$ line.
We thus use these 16762 AGN with good quality of the Ly$\alpha$ line to compose our AGN sample, referred to as All-AGN sample. 

To investigate the type dependence, we classify these 16762 AGN into type-$1$ and type-$2$ AGN. In the same manner as the T1-AGN(H) sample construction, we use Ly$\alpha$ line width measurements of \cite{Rakshit20} for the type-1 and type-2 AGN classification.
For the 16762 AGN, we apply the criterion of Ly$\alpha$ FWHM $> 1000$ km s$^{-1}$ \citep{Villarroel+14,Panessa+02} to select type-1 AGN, and obtain 14693 type-1 AGN. Following \cite{Villarroel+14,Panessa+02}, we classify type-2 AGN by the criterion of Ly$\alpha$ FWHM $< 1000$ km s$^{-1}$ and obtain 2069 type-2 AGN \citep[c.f.][]{Alexandroff+13,Zakamska+03}.
These type-1 and type-2 AGN are referred to as T1-AGN and T2-AGN, respectively.

Table \ref{tab:fg} presents the summary of foreground samples. We obtain 14693 and 1829 type-1 AGN, which referred to as T1-AGN and T1-AGN(H), from the SDSS and HETDEX surveys, respectively. We select 2069 type-2 AGN that are referred to as T2-AGN from the SDSS survey. 

\subsection{Background Source Sample} \label{subsec:bkagn}

In this subsection, we describe how the background sources are selected. We select the background sources with the SDSS DR16Q catalog, following the three steps below.

In the first step, we extract QSOs in our study field from the SDSS DR16Q catalog. We then select QSOs falling in the range of redshifts from 2.08 to 3.67.
The lower and upper limits of the redshift range are determined by the Ly$\alpha$ forest.
Our goal is to probe {\sc Hi} absorbers at $z=2.0-3.0$ with the Ly$\alpha$ forest.
Because the Ly$\alpha$ forest is observed in the rest-frame $1040-1185$ \AA\ of the background sources, we obtain the lower and upper limits of the redshifts, 2.08 and 3.67, by 
$1216\times (1+2.0)/1185-1=2.08$ and $1216\times(1+3.0)/1040-1=3.67$, respectively.
By this step, we have selected 26899 background source candidates.

In the second step, we choose background source candidates with good quality.
We calculate the average signal to noise ratio, $\langle$S/N$\rangle$, in the wavelength range of the Ly$\alpha$ forest for the 26899 background source candidates, and select 15573 candidates with $\langle$S/N$\rangle$ greater than 1.4. 
To maximize the special resolution of the tomography map, we set the threshold, $\langle$S/N$\rangle$ $>1.4$, smaller than the value used by \cite{mukae+20}.
This threshold is more conservative than the value, 1.2, used in \cite{Lee+18}.
In the third step, we remove damped Ly$\alpha$ absorbers (DLAs) and broad absorption lines (BALs) from the Ly$\alpha$ forest of the 15573 candidates, because the DLAs and BALs cause an overestimation of the absorption of the Ly$\alpha$ forest. 
We identify and remove DLAs using the catalog of \cite{Chabanier+22}, which is based on the SDSS DR16Q \citep{Lyke+20}.
We mask out the wavelength ranges contaminated by the DLAs of the \cite{Chabanier+22} catalog (see Section \ref{subsec:masking} for the procedures). We conduct visual inspection for the 15573 candidates to remove 115 BALs.
In this way, we obtain 15458 ($=15573-115$) sources whose spectra are free from DLAs and BALs, which we refer to as the background source sample. Table \ref{tab:bg} lists the number of background sources in each field.


\section{HI Tomography and Mapping} \label{sec:hi_tomography}

In this section we describe the process to construct {\sc Hi} tomography maps with the spectra of the background sources. For {\sc Hi} tomography, we need to obtain intrinsic continua of the background sources. Section \ref{subsec:fitting} explains masking the biasing absorption features in the background sources, while Section \ref{subsec:reconstruction} determines the intrinsic continua of the background source spectra. In Section \ref{subsec:reconstruction}, we construct {\sc Hi} tomography maps with the intrinsic continuum spectra.

\subsection{DLA and Intrinsic Absorption Masking} \label{subsec:masking}

Because a DLA is an absorption system with a high neutral hydrogen column density $N_\mathrm{\sc HI} > 2 \times 10^{20}$ cm$^{-2}$, the intervening DLA completely absorbs a large portion of the Ly$\alpha$ forest over $\Delta$v $\sim$ 10$^3$ km s$^{-1}$, which gives bias in the estimates of the intrinsic continua of the background sources.
For the spectra of the background sources, we mask out the DLAs identified in Section \ref{subsec:bkagn}. 
We determine the range of wavelengths for masking with the IDL code of \cite{Lee+12}.
The wavelength range corresponds to the equivalent width of each DLA \citep{Draine+11}:
\begin{equation} \label{eq:dla}
    W \sim \lambda_\mathrm{\alpha}\left[\frac{e^2}{m_\mathrm{e}c^2}N_\mathrm{HI} f_\mathrm{\alpha}\lambda_\mathrm{\alpha}\left(\frac{\gamma_\mathrm{\alpha} \lambda_\mathrm{\alpha}}{c}\right)\right]^{1/2}.
\end{equation}
In the formula, $\lambda_{\alpha}$ is the rest-frame wavelength of the hydrogen Ly$\alpha$ line (i.e. 1216 \AA),
while $c$, $e$, $m_e$, $f_\alpha$, $N_{\sc Hi}$, and $\gamma_\alpha$ are the speed of light, the electron charge, the electron mass, the Ly$\alpha$ oscillator strength, the {\sc Hi} column density of the DLA, and the sum of the Einstein A coefficients.
We mask out these wavelength ranges of the background source spectra.
In Figure \ref{fig:MFPCA}, the masked DLA is indicated by yellow hatches.

We also mask out the intrinsic absorption lines of the metal absorption lines, which are the other sources of bias.
We mask {\sc SIv} $\lambda$1062, {\sc Nii} $\lambda$1084, {\sc Ni} $\lambda$1134, and {\sc Ciii} $\lambda$1176 \citep{Lee+12}, which are shown by the dashed lines in Figure \ref{fig:MFPCA}.
Because the spectral resolutions of SDSS DR14Q are $\Delta\lambda=1.8-5.2$ \AA, we adopt the masking size of $10$ \AA\ in the observed frame.

\subsection{Intrinsic Continuum Determination} \label{subsec:fitting}

In order to obtain the intrinsic continuum of the background source (Section \ref{subsec:bkagn}) in the Ly$\alpha$ forest
wavelength range (rest-frame $1040-1185$ \AA), we conduct mean-flux regulated principle component analysis (MF-PCA) fitting with the IDL code \citep{Lee+12} for the background sources after the masking (Section \ref{subsec:masking}).

There are two steps in the MF-PCA fitting process.
The first step is to predict the shape of the intrinsic continuum of the background sources in the Ly$\alpha$ forest wavelength range. We conduct least-squares principle component analysis (PCA) fitting \citep{suzuki+05,Lee+12} to the background source spectrum in the rest frame $1216 - 1600$ \AA:
\begin{equation} \label{eq:pca}
    f_{\rm PCA} (\lambda) = \mu(\lambda)+\sum_{j=1}^{8} c_j\xi_j (\lambda),
\end{equation}
where $\lambda$ is the rest-frame wavelength. The values of $c_j$ are the free parameters for the weights.
The function of $\mu(\lambda)$ is the average spectrum calculated from the 50 local QSO spectra in \cite{suzuki+05}.
The function of $\xi_j(\lambda)$ represents the $j$th principle component (or ‘eigenspectrum’) out of the 8 principle components taken from the PCA template derived \cite{suzuki+05}.

\begin{figure}[ht!]
\begin{center}
\includegraphics[scale=0.22]{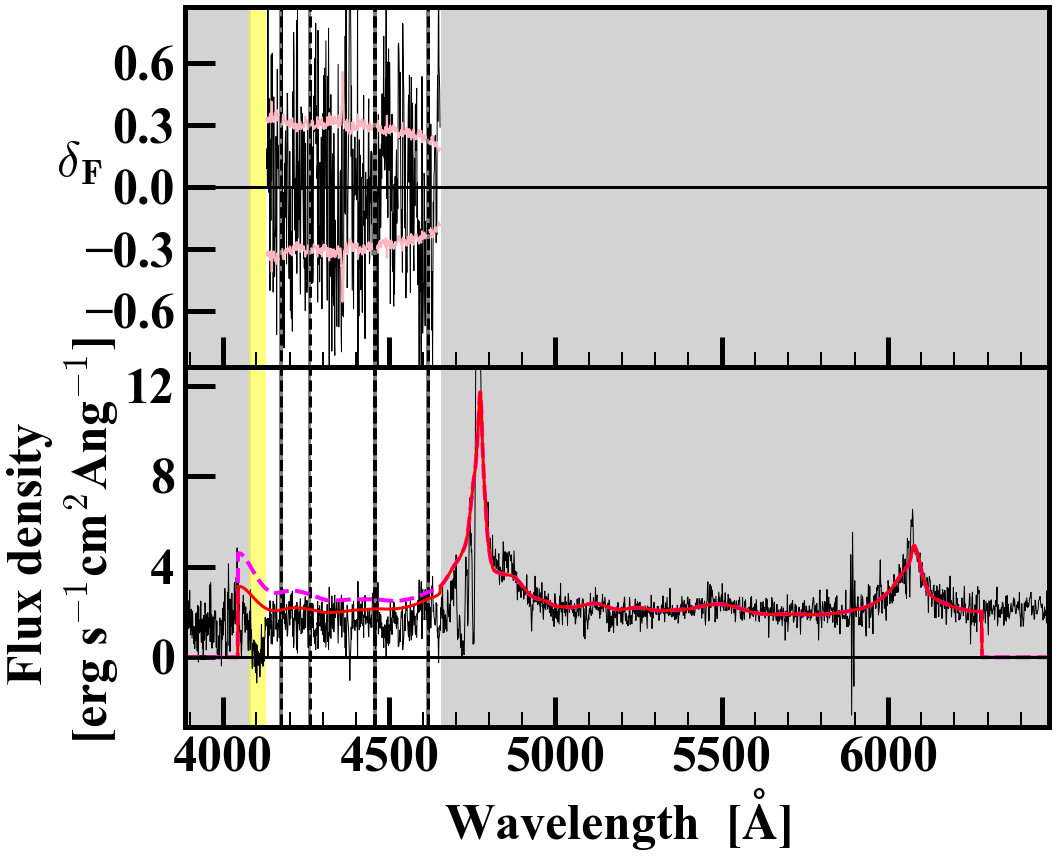}
\end{center}
\caption{Example of a background source spectrum that was used for the reconstruction of the {\sc Hi} tomography map. Bottom panel: Estimation of intrinsic continuum. The thin black line is the spectrum of a background source taken from the SDSS survey. The red and magenta lines are the results of MF-PCA and PCA fitting, respectively.
The vertical dashed lines present the central wavelengths of the metal absorptions. 
The gray hatches represent the wavelength ranges that are not used for the {\sc Hi} tomography map reconstructions.
The yellow hatch indicates the wavelength ranges of DLA. Top panel: Spectrum of $\delta_\mathrm{F}$ extracted from the bottom panel in the Ly$\alpha$ forest wavelength range. The vertical yellow and gray hatches are the same as those in the bottom panel. The black and pink lines show the spectrum of $\delta_\mathrm{F}$ and the error of $\delta_\mathrm{F}$ at the corresponding wavelength extracted from the bottom panel. The horizontal line indicates the cosmic average of Ly$\alpha$ forest transmission.}
\label{fig:MFPCA}
\end{figure}

In the second step, we predict the intrinsic continuum of the background source in the Ly$\alpha$ forest wavelength range.
Because the PCA template is obtained with the local QSO spectra, the best-fit $f_{\rm PCA}$ in the Ly$\alpha$ forest does not include cosmic evolution on the average transmission rate.
On average, the best-fit $f_{\rm PCA}$ in the Ly$\alpha$ forest should agree with the cosmic mean-flux evolution \citep{FG+08}:
\begin{equation} \label{eq:mf}
    \langle F(z) \rangle  = \exp[-0.001845(1 + z)^{3.924}],
\end{equation}
where $z$ is the redshift of the absorber.
We use $f_{\rm PCA}$ and a correction function of $a+b\lambda$ to estimate the intrinsic continuum $f_{\rm intrinsic}(\lambda)$ for large-scale power along the line of sight with the equation:
\begin{equation} \label{eq:mf1}
    f_{\rm intrinsic}(\lambda) = f_{\rm PCA  }(\lambda)\times(a+b\lambda),
\end{equation}
where $a$ and $b$ are the free parameters.
Because the ratio of $f_{\rm obs} (\lambda)/f_{\rm intrinsic}(\lambda)$ should agree with the cosmic average $\langle F(z) \rangle $
for $z = (\lambda/1216) - 1$ in the wavelength range of the Ly$\alpha$ forest, we conduct least-squares-fitting to find the values of $a$ and $b$ providing the best fit between the mean ratio and the cosmic average.
The red line shown by the bottom panel of Figure \ref{fig:MFPCA} presents a MF-PCA fitted continuum derived from the spectrum of one of our background sources.

\begin{figure}[ht!]
\begin{center}
\includegraphics[scale=0.22]{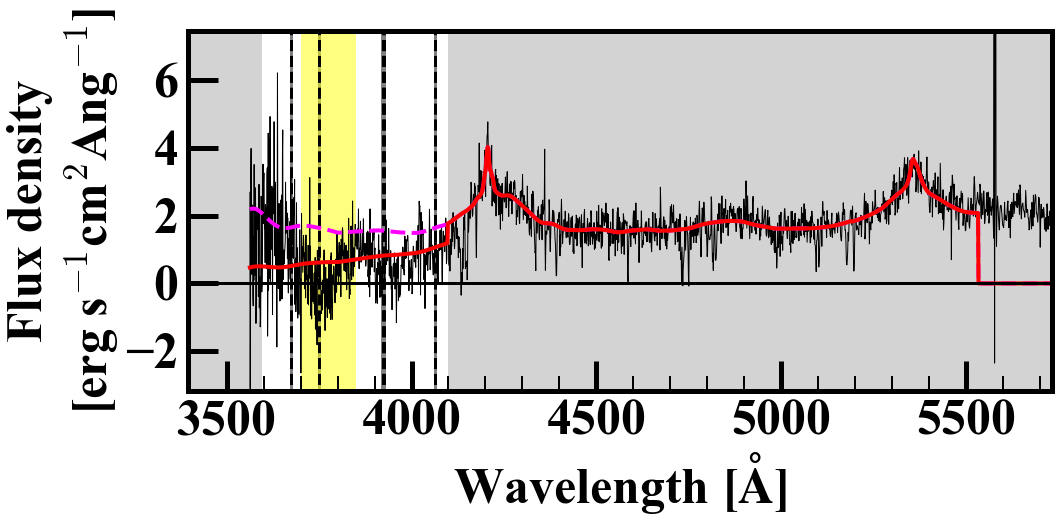}
\end{center}
\caption{Same as the bottom panel of Figure \ref{fig:MFPCA}, but for the background spectrum with a poor fitting result. The red and magenta lines are the results of MF-PCA and PCA continuum fitting, respectively. The yellow hatch indicates the wavelength range of unknown absorption.}
\label{fig:Poor_fitting}
\end{figure}

By the MF-PCA fitting, we have obtained the estimates of $f_{\rm intrinsic}(\lambda)$ for 14736 out of the 15458 background sources. We find the other background sources show poor fitting results found by visual inspection.
We do not use these background sources in the following analyses.
Figure \ref{fig:Poor_fitting} shows an example of poor fitting result due to the unknown absorption.
We adopt continuum fitting errors of $\sim7\%, \sim6\%$, and $\sim4\%$ for Ly$\alpha$ forests with mean S/N values of $<4$, $4-10$, and $>10$, respectively \citep{Lee+12}.


\subsection{HI Tomography Map Reconstruction} \label{subsec:reconstruction}
We reconstruct our {\sc Hi} tomography maps by a procedure similar to \cite{Lee+18}.
We define Ly$\alpha$ forest fluctuations $\delta_{F}$ at each pixel on the spectrum by
\begin{equation} \label{eq:df}
    \delta_{\rm F} = \frac{f_{\rm obs} / f_{\rm intrinsic}} {\langle F(z) \rangle } - 1,
\end{equation}
where $f_{obs}$ and $f_{intrinsic}$ are the observed spectrum and estimated intrinsic continuum, respectively. ${\langle F(z) \rangle }$ is the cosmic average transmission.
We calculate $\delta_{\rm F}$ with our background source spectra.
The top panel of Figure \ref{fig:MFPCA} shows the `spectrum' of $\delta_{\rm F}$ derived from the $f_{obs}$ and $f_{intrinsic}$ in the bottom panel.
For the pixels in the wavelength ranges of masking (Section \ref{subsec:masking}), we do not use $\delta_{\rm F}$ in our further analyses. We thus obtain $\delta_{\rm F}$ in 876,560 pixels.

For the the HI tomography map of the Extended Fall field, we define the cells of the {\sc Hi} tomography map in the three-dimensional comoving space.
We choose a volume of $30^{\circ} \times 3.3^{\circ}$ in the longitudinal and latitudinal dimensions, respectively, in the redshift range of $2.0 < z < 3.0$. 
The comoving size of our {\sc Hi} tomography map is 2257 $h^{-1} {\rm cMpc}$ $\times$ 233 $h^{-1} {\rm cMpc}$ $\times$ 811 $h^{-1} {\rm cMpc}$ in the right ascension (R.A.), declination (Dec), and $z$ directions, respectively in the same manner as \cite{mukae+20}.
Our {\sc Hi} tomography map has $451 \times 46 \times 162$ cells, 
and one cell is a cubic with a size of 5.0 $h^{-1} {\rm cMpc}$ on a side, where the line-of-sight 
distance is estimated under the assumption of the Hubble flow.

\begin{figure*}[ht!]
\begin{center}
\includegraphics[scale=0.2]{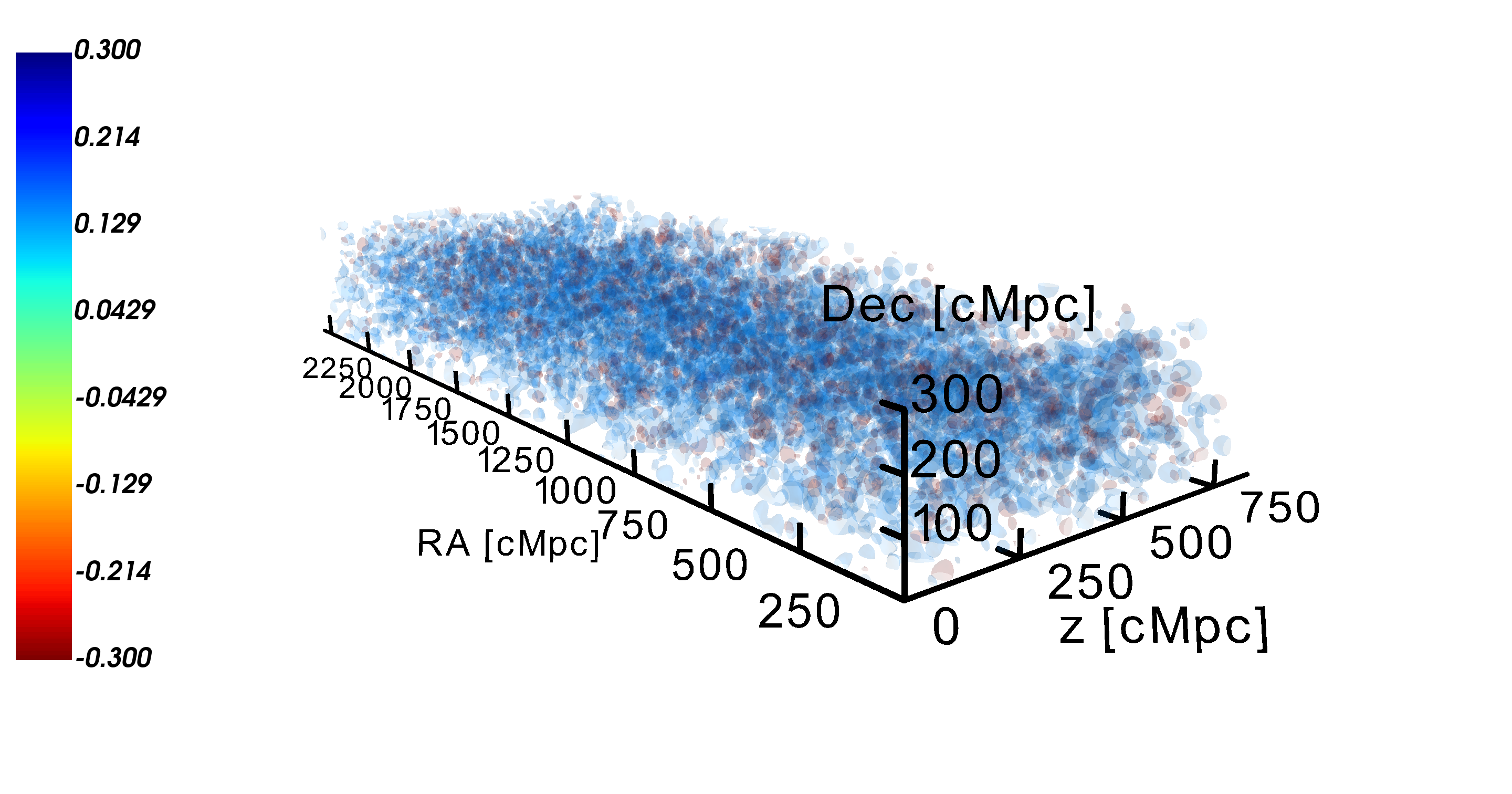}
\end{center}
\caption{3D {\sc Hi} tomography map of the ExFall field. The color
contours represent the values of $\delta_{\rm F}$ from negative (red) to positive (blue).
The spatial volume of the {\sc Hi} tomography map is $2257 \times 233 \times 811$ $h^{\rm -3}$cMpc$^3$.
The redshift range is $z = 2.0 - 3.0$.}
\label{fig:tomography_map_fall}
\end{figure*}

We conduct a Wiener filtering scheme for reconstructing the sightlines that do not have background sources. 
We use the calculation code developed by \cite{stark+15a}.
The solution for each cell of the reconstructed sightline
is obtained by
\begin{equation} \label{eq:wf}
    \delta_{\rm F}^{\rm rec} = \mathrm{C}_{\mathrm{MD}}\cdot(\mathrm{C}_{\mathrm{DD}}+\mathrm{N})^{-1}\cdot\delta_{\rm F},
\end{equation}
where $\mathrm{C}_{\mathrm{MD}}$, $\mathrm{C}_{\mathrm{DD}}$, and $\mathrm{N}$ are the map-data, data-data, and noise covariances, respectively.
We assume Gaussian covariances between two points $\mathrm{r}_1$ and $\mathrm{r}_2$:
\begin{equation} \label{eq:covariance}
    \mathrm{C}_{\mathrm{MD}} = \mathrm{C}_{\mathrm{DD}} = \mathrm{C}(\mathrm{r}_\mathrm{1},\mathrm{r}_\mathrm{2}),
\end{equation}
\begin{equation} \label{eq:gaussian covariance}
    \mathrm{C}(\mathrm{r}_\mathrm{1},\mathrm{r}_\mathrm{2}) = \sigma^{2}_{F}\exp \left[ -\frac{(\Delta r_{\|})^2}{2L^2_{\|}} \right] \exp \left[ -\frac{(\Delta r_{\bot})^2}{2L^2_{\bot}} \right],
\end{equation}
where $\Delta r_{\|}$ and $\Delta r_{\bot}$ are the distances between $\mathrm{r}_\mathrm{1}$ and $\mathrm{r}_\mathrm{2}$ in the directions of parallel and transverse to the line of sight, respectively. 
The values of $L_{\bot}$ and $L_{\|}$ are the correlation lengths for vertical and parallel to the line-of-sight (LoS) direction, respectively, and defined with $L_{\bot}$ = $L_{\|}$ = 15 $h^{-1} {\rm cMpc}$.
The value of $\sigma^2_F$ is the normalization factor that is 
$\sigma^2_F = 0.05$.
\cite{stark+15a} develop this Gaussian form to obtain a reasonable estimate of the true correlation function of the Ly$\alpha$ forest. 
We perform the Wiener filtering reconstruction with the values of $\delta_{\rm F}$ at the 898390 pixels, using 
the aforementioned parameters of the \cite{stark+15a} algorithm with a stopping tolerance of $10^{-3}$ for the pre-conditioned conjugation gradient solver.
As noted by \cite{Lee+16}, the boundary effect that leads to an additional error on $\delta_{\rm F}$ occurs at the positions that are near the boundaries of an {\sc Hi} tomography map.
The boundary effect is caused by the background sightlines not covering the region that contribute to the calculation of the $\delta_{F}$ values for cells near the {\sc Hi} tomography map boundaries.
To avoid the boundary effect, we extend a distance of 40 h$^{-1}$cMpc for each side of the {\sc Hi} tomography map of the ExFall field.
The resulting map is shown in Figure \ref{fig:tomography_map_fall}.

For the HI tomography map reconstruction of the Extended Spring field (hereafter ExSpring field), we perform almost the same procedure as the one of the ExFall field.
The area of the ExSpring field is more than 6 times larger than that of the ExFall field.
We separate the ExSpring field into $8\times3=24$ footprints to save calculation time.
Each footprint covers an area of $10^\circ \times 5^\circ$ in the R.A. and Dec directions, respectively.
We reconstruct the {\sc Hi} tomography map one by one for the footprints of the ExSpring field.

\begin{figure*}[ht!]
\begin{center}
\includegraphics[scale=0.2]{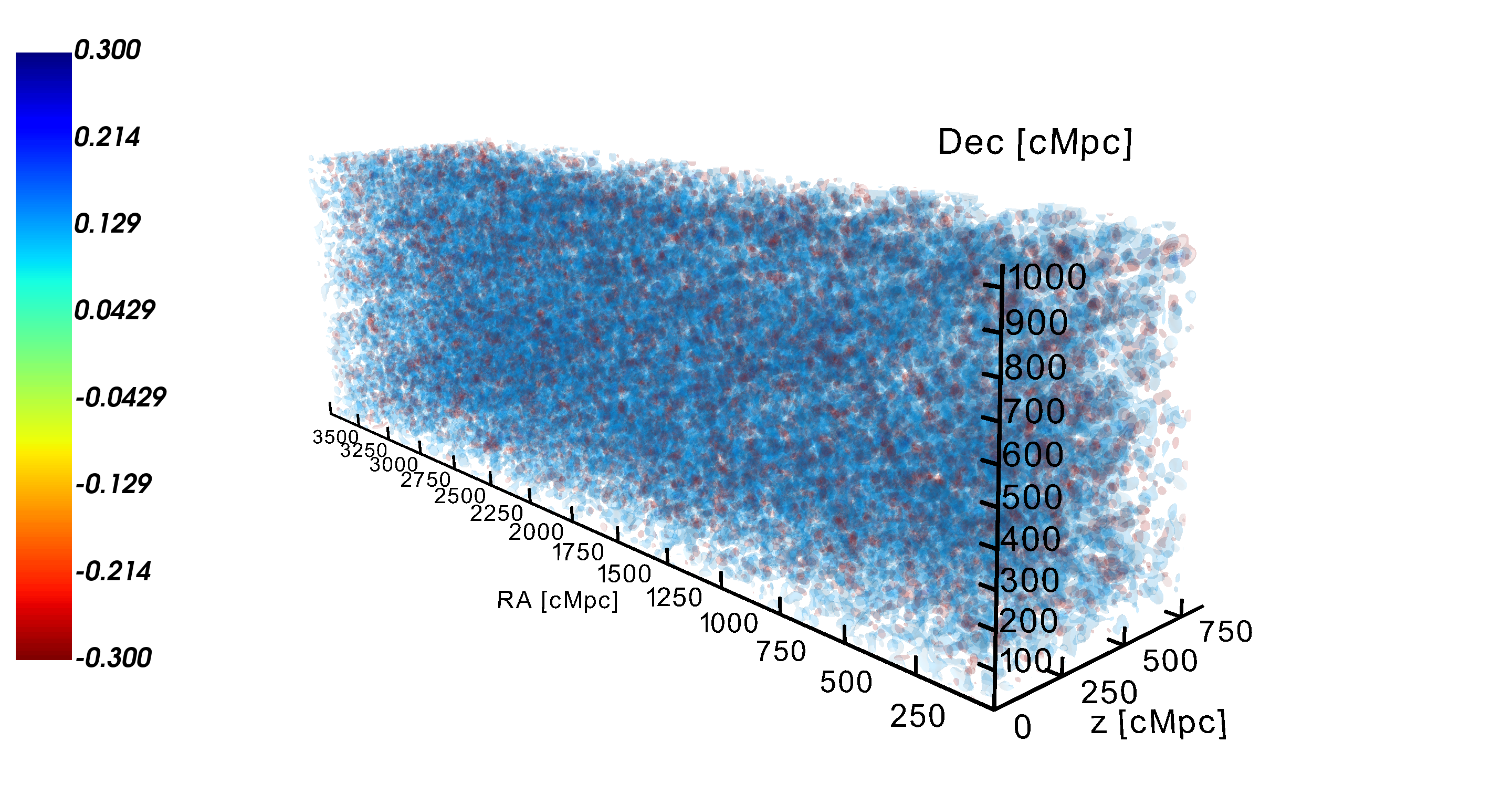}
\end{center}
\caption{Same as Figure \ref{fig:tomography_map_fall}, but for the ExSpring field.
The spatial volume of the {\sc Hi} tomography map is $3475 \times 1058 \times 811$ $h^{\rm -3}$cMpc$^3$.
}
\label{fig:tomography_map_spring}
\end{figure*}

To weaken the boundary effect, we extend a distance of 40 h$^{-1}$cMpc for each side of the footprints.
The extensions mean that every two adjacent footprints has an overlapping region of 80 h$^{-1}$cMpc width.
The width of the overlapping regions is a conservative value to weaken the boundary effect since it is much larger than the resolution, 15 h$^{-1}$cMpc, of our {\sc Hi} tomography maps.
By the 40 h$^{-1}$cMpc extension, we reduce the uncertainty in the $\delta_{F}$ value for the edge of each footprint caused by boundary effect to $\pm$0.01.
This value corresponds to the $1/10$ of the typical error for each cell of the {\sc Hi} tomography map \citep{mukae+20}
The remaining additional error caused by boundary effect is negligible compared to the statistical uncertainties in the HI distributions obtained in Section \ref{sec:result}.
Then we follow the reconstruction procedure for the ExFall field to reconstruct HI tomography maps of the footprints and cut off all the cells within 40 h$^{-1}$cMpc to the borders that are affected by the boundary effect.
Finally we obtain the {\sc Hi} tomography map of the ExSpring field with a special volume of 3475 $h^{-1} {\rm cMpc}$ $\times$ 1058 $h^{-1} {\rm cMpc}$ $\times$ 811 $h^{-1} {\rm cMpc}$ in the R.A., Dec, and $z$ directions, respectively (Figure \ref{fig:tomography_map_spring}).

\section{Results and Discussions} \label{sec:result}

\subsection{Average HI Profiles around AGN: Validations of our AGN Samples} 
\label{subsec:average_profile}

In this section we present the {\sc Hi} profile, $\delta_\mathrm{F}$ as a function of distance, with the All-AGN sample sources, using the reconstructed {\sc Hi} tomography maps.
We compare the {\sc Hi} profile of the All-AGN sample to the one of the previous study \citep{FR+13}.
We also present the comparison of the {\sc Hi} profiles between T1-AGN(H) and T1-AGN samples that are made with the HETDEX and SDSS data.
In this study, we only discuss the structures having size $\gtrsim 15$ $h^{-1}$cMpc corresponding to the resolution of our 3D {\sc Hi} tomography maps.

For the {\sc Hi} profiles with the All-AGN sample, we extract $\delta_\mathrm{F}$ values around the 16978 All-AGN sample sources in the {\sc Hi} tomography map.
We cut the {\sc Hi} tomography map centered at the positions of the All-AGN sample sources, and stack the $\delta_\mathrm{F}$ values to make a two dimensional (2D) map of the average $\delta_\mathrm{F}$ distribution around the sources that is referred to as a 2D {\sc Hi} profile of the All-AGN sample sources. The two dimensions of the 2D {\sc Hi} profile correspond to the transverse distance $D_\mathrm{Trans}$ and the LoS Hubble distance. The velocity corresponding to the LoS Hubble distance is referred to as the LoS velocity.

\begin{figure*}[ht!]
\begin{center}
\includegraphics[scale=1]{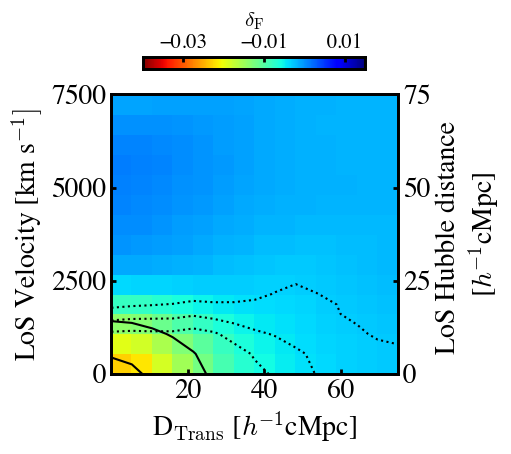}
\end{center}
\caption{
2D {\sc Hi} profile of the All-AGN sample sources. The color map indicates the $\delta_\mathrm{F}$ values of each cell of the 2D {\sc Hi} profile.
The solid lines denote constant $\delta_\mathrm{F}$ values in steps of $-0.01$ starting at $-0.01$. The dotted lines correspond to multiples of $3\sigma$ starting at $6\sigma$. }
\label{fig:all_agn_2d}
\end{figure*}

Figure \ref{fig:all_agn_2d} shows the 2D {\sc Hi} profile with values of $\delta_{\rm F}$ for All-AGN sample.
The solid black lines denote the contours of $\delta_{\rm F}$. 
In each cell of the 2D {\sc Hi} profile, we define the $1\sigma$ error with the standard deviation of $\delta_{\rm F}$ values of the 100 mock 2D {\sc Hi} profiles.
Each mock 2D {\sc Hi} profile is obtained in the same manner as the real 2D {\sc Hi} profile, but with random positions of sources whose number is the same as the one of All-AGN sample sources.
In Figure \ref{fig:all_agn_2d}, the dotted black lines indicate the contours of the 6$\sigma$, 9$\sigma$ and $12\sigma$ confidence levels, respectively.
We find the $19.5\sigma$ level detection of $\delta_{\rm F}$ at the source position (0,0).
The $\delta_{\rm F}$ value at the source position indicates the averaging value over the ranges of ($-7.5$ $h^{-1}\mathrm{cMpc}$, $+7.5$ $h^{-1}\mathrm{cMpc}$) in both the LoS and transverse directions.
The $19.5\sigma$ level detection at the source position is suggestive that obvious Ly$\alpha$ forest absorption exists near the All-AGN sources on average
The 2D {\sc Hi} profile is more extended in the transverse direction than along the line of sight.
We discuss this difference in Section \ref{subsec:AGN_LoSandTrans_profile}.

\begin{figure}[ht!]
\begin{center}
\includegraphics[scale=0.48]{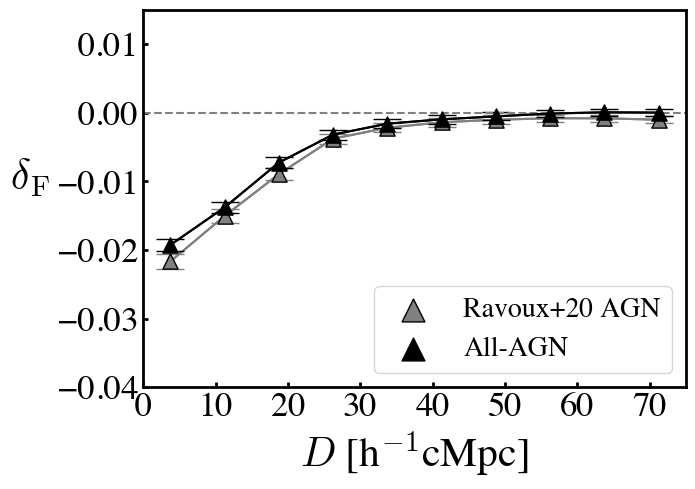}
\end{center}
\caption{
{\sc Hi} radial profile of the All-AGN and \cite{ravoux+20} AGN samples.
The black and gray data points and error bars show the {\sc Hi} radial profiles of our All-AGN sample sources and the AGN of \citealt{ravoux+20}, respectively. The horizontal dashed line shows the cosmic average Ly$\alpha$ transmission fluctuation, $\delta_\mathrm{F}=0$.}
\label{fig:ravoux+20}
\end{figure}

We then define a 3D distance, $D$, under the assumption of the Hubble flow in the LoS direction. We derive $\delta_{\rm F}$ as a function of $D$ that is referred to as "{\sc Hi} radial profile", averaging $\delta_{\rm F}$ values of the 2D {\sc Hi} profile over the 3D distance.
Figure \ref{fig:ravoux+20} shows the {\sc Hi} radial profile of the All-AGN sample.
We find that the $\delta_{\rm F}$ values increase towards a large distance.
This trend is consistent with the one found by \cite{ravoux+20} with the SDSS quasars.

\cite{ravoux+20} have obtained the average Ly$\alpha$ transmission fluctuation distribution around the AGN taken from the SDSS data release 16 quasar (SDSS DR16Q) catalog in the field of Strip 82. The criteria of the target selection for the SDSS DR16Q and SDSS DR14Q sources are the same. The luminosity distribution of AGN for \cite{ravoux+20} is almost the same as that of our All-AGN sample sources that are taken from the SDSS DR14Q catalog.
We derive the average radial {\sc Hi} profile of the \cite{ravoux+20} AGN sources by the same method as for our All-AGN sample, using the 3D {\sc Hi} tomography map reconstructed by \cite{ravoux+20}.
We compare the radial {\sc Hi} profile of the All-AGN sample with the one derived from the 3D {\sc Hi} tomography map of \cite{ravoux+20}.
The comparison is shown in Figure \ref{fig:ravoux+20}.
Our result agrees with that of \cite{ravoux+20} within the error range at scale $D> 10$ $h^{-1}$ cMpc.
The peak values of $\delta_{\rm F}$ showing the strongest Ly$\alpha$ absorption are comparable, $\delta_{\rm F}\simeq -0.02$. 
The slight difference between the peak values of our and \citeauthor{ravoux+20}'s results can be explained by the different approaches of the estimation for the intrinsic continuum adopted by \citeauthor{ravoux+20} and us.
\citeauthor{ravoux+20} conduct power law fitting, which is different from the MF-PCA fitting that we used, for the intrinsic continuum in the wavelength range of the Ly$\alpha$ forest.
Given the low ($\sim 15$ $h^{-1}$) spatial resolution of both our {\sc Hi} tomography map and that of \cite{ravoux+20}, neither studies are able to search for the proximity effect making a photoionization region around AGN \citep{D'Odorico+08}.
From the comparison shown by Figure \ref{fig:ravoux+20}, we conclude that the Ly$\alpha$ forest absorption derived from our {\sc Hi} tomography map is reliable.

To check the reliability of the HETDEX survey results, we use the reliable result of the SDSS AGN to compare with the result derived by the HETDEX AGN.

\begin{figure}[ht!]
\begin{center}
\includegraphics[scale=0.6]{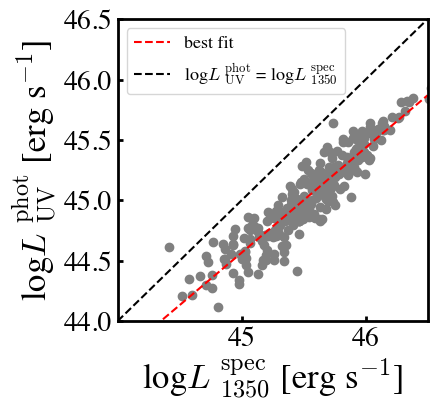}
\end{center}
\caption{Relations of $L^\mathrm{phot}_\mathrm{UV}$ against $L^\mathrm{spec}_\mathrm{1350}$ for the sources both categorized in the T1-AGN(H) and T1-AGN samples. The $L^\mathrm{phot}_\mathrm{UV}$ and $L^\mathrm{spec}_\mathrm{1350}$ are measured from the HSC r-band imaging and SDSS spectra \citep{Rakshit20}, respectively. The gray points show the distribution of $L^\mathrm{spec}_\mathrm{1350}$ $-$ $L^\mathrm{phot}_\mathrm{UV}$ relations for the sources both categorized in the T1-AGN(H) and T1-AGN samples. The black dashed line indicates the relation where $L^\mathrm{spec}_\mathrm{1350}$ $=$ $L^\mathrm{phot}_\mathrm{UV}$. The red dashed line represents the linear best fit of the blue points.}
\label{fig:luv_vs_l1350}
\end{figure}

We select type-1 AGN from the HETDEX's T1-AGN(H) and SDSS's T1-AGN samples to make sub-samples of T1-AGN(H) and T1-AGN with matching rest-frame 1350 \AA \ luminosity ($L_\mathrm{1350}$).
For T1-AGN, the measurements directly from the SDSS spectra ($L^\mathrm{spec}_\mathrm{1350}$) are available \citep{Rakshit20}. For T1-AGN(H), we do not have $L^\mathrm{spec}_\mathrm{1350}$ measurements from the HETDEX spectra, we estimate it using HSC r-band imaging. Since the central wavelength of the r-band imaging is rest-frame $\sim1700 {\rm \AA}$, we calibrate the conversion between r-band luminosity, $L^\mathrm{phot}_\mathrm{UV}$, and $L^\mathrm{spec}_\mathrm{1350}$. We examine the 283 type-1 AGN sources that appear in both the SDSS and HETDEX surveys (and, thus, have both $L^\mathrm{spec}_\mathrm{1350}$ measurements from SDSS and r-band luminosities from HSC) to calibrate the relationship. The results are displayed in Figure \ref{fig:luv_vs_l1350}.
The $L^\mathrm{phot}_\mathrm{UV}$ are always smaller than those of $L^\mathrm{spec}_\mathrm{1350}$ \citep{Rakshit20}.
Due to the blue UV slope of the spectra for the AGN both categorized in the T1-AGN(H) and T1-AGN samples, the luminosity of the rest-frame 1350 \AA\ always shows a larger value than the one of rest-frame 1700 \AA.
We conduct linear fitting to the data points of Figure \ref{fig:luv_vs_l1350}, and obtain the best-fit linear function.
With the best-fit linear function, we estimate $L^\mathrm{spec}_\mathrm{1350}$ values for the HETDEX's T1-AGN(H) sample sources.

\begin{figure}[ht!]
\begin{center}
\includegraphics[scale=0.6]{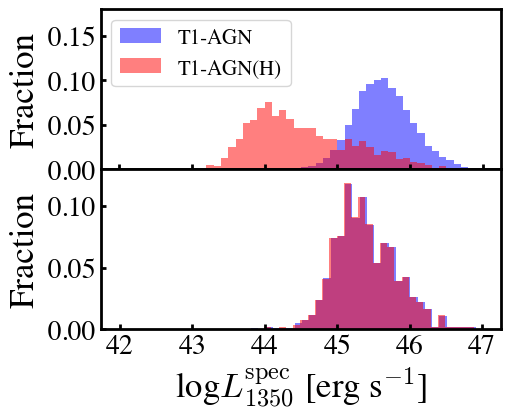}
\end{center}
\caption{Top panel: L$^\mathrm{spec}_\mathrm{1350}$ distributions of the T1-AGN and T1-AGN(H) samples with blue and red histograms, respectively. 
Bottom panel: Same as the top panel, but for the T1-AGN and T1-AGN(H) sub-sample sources.}
\label{fig:pdf_sdsst1agn_hetdext1agn}
\end{figure}

We show the $L^\mathrm{spec}_\mathrm{1350}$ distributions of all the T1-AGN(H) and T1-AGN sample sources in the upper panel of Figure \ref{fig:pdf_sdsst1agn_hetdext1agn}.
The $L^\mathrm{spec}_\mathrm{1350}$ distribution of T1-AGN(H) covers a wider luminosity range than the one of T1-AGN. To make sure the comparison between the SDSS and HETDEX AGN is fair,
we make the sub-samples of T1-AGN and T1-AGN(H) that consist of the sources with matching $L^\mathrm{spec}_\mathrm{1350}$ distributions.
We present the $L^\mathrm{spec}_\mathrm{1350}$ distributions of the T1-AGN and T1-AGN(H) sub-samples in the bottom panel of Figure \ref{fig:pdf_sdsst1agn_hetdext1agn}.
We obtain 540 and 4338 type-1 AGN for the sub-samples of T1-AGN(H) and T1-AGN, respectively, whose $L^\mathrm{spec}_\mathrm{1350}$ distributions are shown in the bottom panel of Figure \ref{fig:luv_vs_l1350}.

\begin{figure}[ht!]
\begin{center}
\includegraphics[scale=0.48]{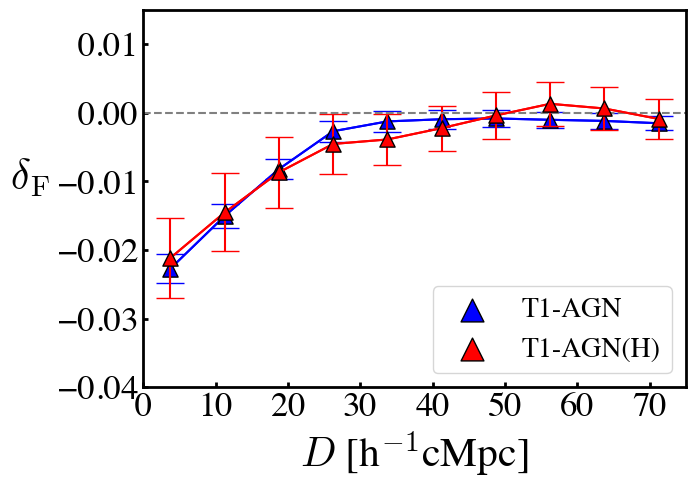}
\end{center}
\caption{{\sc Hi} radial profiles of the T1-AGN and T1-AGN(H) sub-samples. 
The blue and red triangles show the values of $\delta_{\rm F}$ as a function of distance, $D$, for the T1-AGN and T1-AGN(H) sample sources, respectively. 
The horizontal dashed line shows the cosmic average Ly$\alpha$ transmission fluctuation, $\delta_{\rm F}=0$.}
\label{fig:hetdext1agn_sdsst1agn}
\end{figure}

We derive the {\sc Hi} radial profiles for the sub-samples of T1-AGN(H) and T1-AGN sample sources, as shown in Figure \ref{fig:hetdext1agn_sdsst1agn}.
The {\sc Hi} radial profiles of T1-AGN(H) and T1-AGN sub-sample sources are in good agreement.

\subsection{AGN Average Line-of-Sight and Transverse {\sc Hi} Profiles} \label{subsec:AGN_LoSandTrans_profile}

Based on the 2D {\sc Hi} profile of the All-AGN sample (Figure \ref{fig:all_agn_2d}), we find that the Ly$\alpha$ forest absorptions of the All-AGN sample sources
are more extended in the transverse direction.
In this section, we present the {\sc Hi} radial profiles of All-AGN sample in the LoS and transverse directions and compare these two {\sc Hi} radial profiles.

\begin{figure}[ht!]
\begin{center}
\includegraphics[scale=0.43]{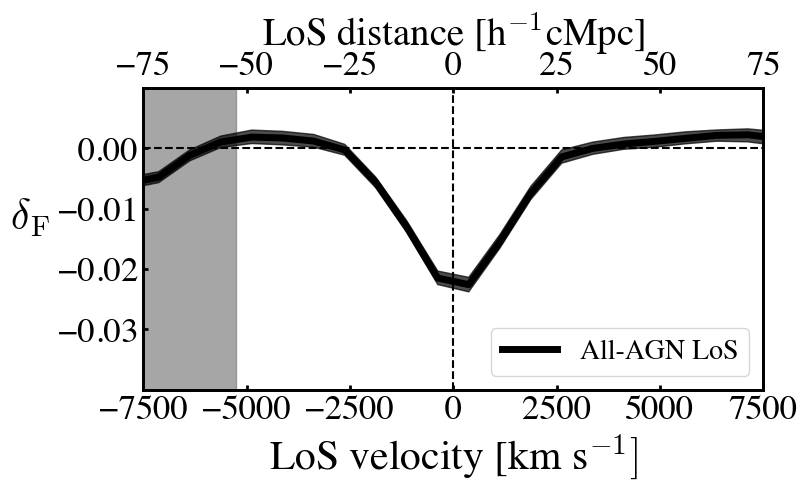}
\end{center}
\caption{{\sc Hi} radial profiles of LoS velocity (LoS distance) for the All-AGN sample. The black solid line shows the $\delta_{\rm F}$ values as a function of LoS velocity (LoS distance) for the All-AGN sample. The vertical dashed line presents the position of LoS velosity $=0$ km s$^{-1}$ (LoS distance $=0$ $h^{-1}$cMpc). The horizontal dashed indicates the cosmic average Ly$\alpha$ transmission fluctuation ,$\delta_{\rm F}=0$. The gray shaded area shows the range of the $\delta_{\rm F}$ not used to derive LoS {\sc Hi} radial profile.}
\label{fig:allagn_con}
\end{figure}

\begin{figure}[ht!]
\begin{center}
\includegraphics[scale=0.48]{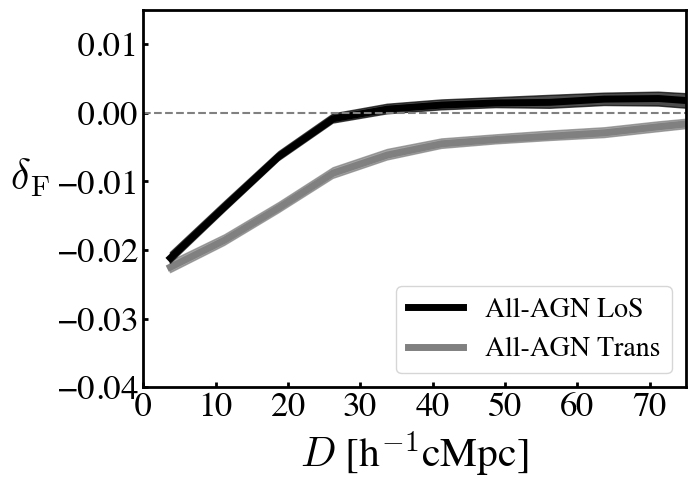}
\end{center}
\caption{LoS and Transverse {\sc Hi} radial profiles the All-AGN sample. The black and gray lines show the $\delta_\mathrm{F}$ values as a function of LoS distance and $D_{\rm Trans}$, respectively. The horizontal dashed line indicates $\delta_\mathrm{F}$=0.}
\label{fig:all_agn_LT}
\end{figure}

To derive the {\sc Hi} radial profile of the All-AGN sample with the absolute LoS distance, which is referred to as the LoS {\sc Hi} radial profile (Figure \ref{fig:all_agn_LT}), we average $\delta_{\rm F}$ values of the 2D {\sc Hi} profiles of All-AGN over $D_\mathrm{Trans}$ $<7.5$ $h^{-1}\mathrm{cMpc}$ (from $-7.5$ $h^{-1}\mathrm{cMpc}$ to $+7.5$ $h^{-1}\mathrm{cMpc}$ in the transverse direction) that corresponds to the spatial resolution of the 2D {\sc Hi} profile map, $15$ $h^{-1}\mathrm{cMpc}$.
Among the 16,978 All-AGN sample sources, 10,884 sources are used as both background and foreground sources. In this case, the Ly$\alpha$ transmission fluctuation ($\delta_{\rm F}$) of these 10,884 sources at the LoS velocity $\lesssim-5250$ km s$^{-1}$ is estimated mainly from their own spectrum. As the discussion in \citet{Youles+22}, the redshift uncertainty of the SDSS AGN causes the overestimation of intrinsic continuum and the underestimation of $\delta_{\rm F}$ around the metal emission lines such as {\sc Ciii} $\lambda$1176. This leads to a systemic error toward negative $\delta_{\rm F}$ in the {\sc Hi} radial profile of LoS velocity (LoS distance) at the LoS velocity $\lesssim5250$ km s$^{-1}$ (Figure \ref{fig:allagn_con}).
The {\sc Hi} radial profile of LoS velocity (LoS distance) is derived by averaging $\delta_{\rm F}$ values over $D_\mathrm{Trans}$ $<7.5$ $h^{-1}\mathrm{cMpc}$ as a function of the negative and positive LoS velocity (LoS distance).
In this study, we only use the values of $\delta_{\rm F}$ at the LoS distance $>-52.5h^{-1}\mathrm{cMpc}$ (LoS velocity $>-5250$ km s$^{-1}$) to derive the LoS {\sc Hi} radial profile of the All-AGN sample (Figure \ref{fig:all_agn_LT}).
The scale, LoS distance $>-52.5h^{-1}\mathrm{cMpc}$ (LoS velocity $>-5250$ km s$^{-1}$), is determined by the maximum wavelength of the Ly$\alpha$ forest we used, the smoothing scale of the Wiener filtering scheme, and the AGN redshift uncertainty, assumed by \cite{Youles+22}.
After removing the $\delta_{\rm F}$ values affected the systemics in the 2D {\sc Hi} profile, we present the LoS {\sc Hi} radial profile of the All-AGN sample in Figure \ref{fig:all_agn_LT}.

We estimate the {\sc Hi} radial profiles of $D_\mathrm{Trans}$, which is referred to as the Transverse {\sc Hi} radial profile,by averaging the $\delta_{\rm F}$ values over the LoS velocity of $(-750,+750)$ km s$^{-1}$ whose velocity width corresponds to $15$ $h^{-1}$ cMpc in the Hubble-flow distance. The {\sc Hi} radial profile of $D_\mathrm{Trans}$ is also shown in Figure \ref{fig:all_agn_LT}.

We compare the LoS and Transverse {\sc Hi} radial profile. 
The $\delta_{\rm F}$ value increase toward large-scale more rapidly in the LoS direction than those in the Transverse direction (Figure \ref{fig:all_agn_LT}).
This difference may be explained by an effect similar to the Kaiser effect \citep{Kaiser+87}, doppler shifts in AGN redshifts caused by the large-scale coherent motions of the gas towards the AGN.
The LoS {\sc Hi} radial profile is positive, $\delta_{\rm F}\sim 0.002\pm0.0008$, at the large scale, $\gtrsim 30$ $h^{-1}$cMpc.
In Section \ref{sec:fr+13}, we discuss the positive $\delta_{\rm F}$ values of LoS {\sc Hi} radial profiles at large scales and compare our observational result to the models of a previous study, \citet{FR+13}.


\subsection{Source Dependences of the AGN Average HI Profiles} \label{subsec:source_dependence}
In this section, we present 2D and {\sc Hi} radial profiles of the AGN sub-samples to investigate how the average {\sc Hi} density depends on luminosity and AGN type.

\subsubsection{AGN Luminosity Dependence} \label{subsubsec:luminosity_dependence}

\begin{figure}[ht!]
\begin{center}
\includegraphics[scale=0.6]{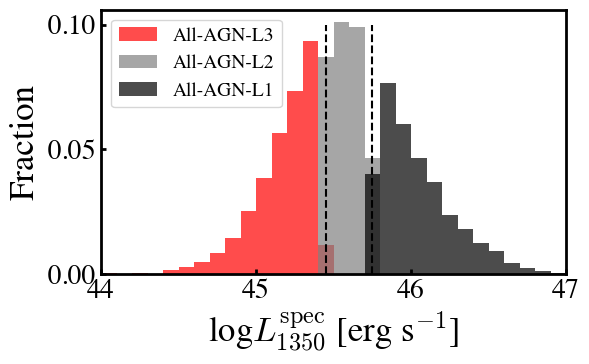}
\end{center}
\caption{log$L^\mathrm{spec}_\mathrm{1350}$ distribution of the bright and All-AGN sample sources. The vertical dashed lines indicate the boarders of $L^\mathrm{spec}_\mathrm{1350}$ where log$(L^\mathrm{spec}_\mathrm{1350}/[{\rm erg \ s ^{-1}}])$ $=45.41$ and $45.75$, respectively.
These three borders separate the All-AGN sample into 3 sub-samples of All-AGN-L3, All-AGN-L2, and All-AGN-L1, respectively.}
\label{fig:bright_faint_t1agn_pdf}
\end{figure}

\begin{figure}[ht!]
\begin{center}
\includegraphics[scale=0.54]{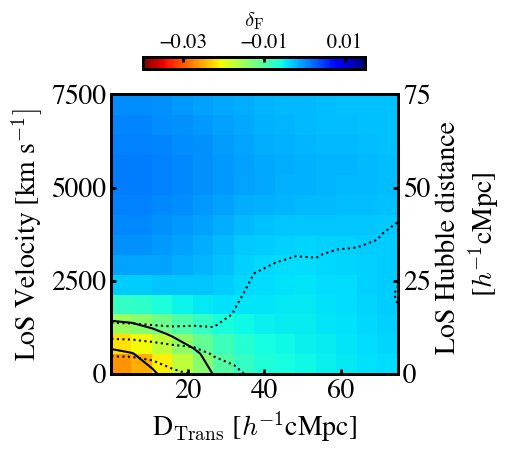}
\includegraphics[scale=0.54]{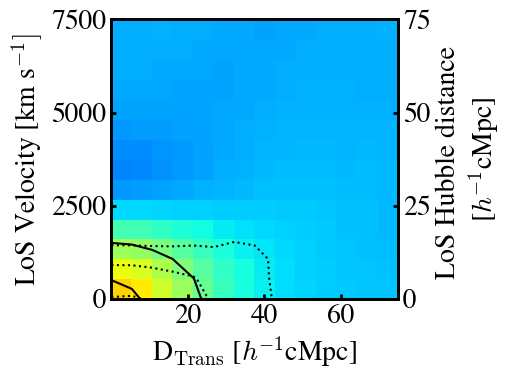}
\includegraphics[scale=0.54]{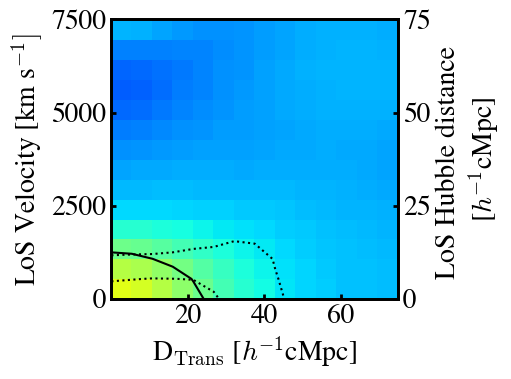}
\end{center}
\caption{Same as Figure \ref{fig:all_agn_2d}, but for the All-AGN-L3 (top), All-AGN-L2 (middle) and All-AGN-L1 (bottom) samples.}
\label{fig:brightandfaint_t1agn_2d}
\end{figure}

We study the AGN-luminosity dependence of the average {\sc Hi} profiles. Figure \ref{fig:bright_faint_t1agn_pdf} presents the $L_{1350}^{\rm spec}$ distribution of All-AGN. We make 3 sub-samples of All-AGN that are All-AGN-L3, All-AGN-L2 and All-AGN-L1.
The luminosity ranges of the sub-samples are $43.70<\log (L_{1350}^{\rm spec}/{\rm [erg\ s^{-1}]})<45.41$, $45.41<\log (L_{1350}^{\rm spec}/{\rm [erg\ s^{-1}]})<45.75$, and $45.75<\log (L_{1350}^{\rm spec}/{\rm [erg\ s^{-1}]})<47.35$, respectively. The luminosity ranges of the 3 sub-samples are defined in a way that the numbers of the AGN are same 5695 in each subsamples.
We derive the 2D {\sc Hi} profiles of the sub-samples in the same manner as Section \ref{subsec:average_profile}, and present the profiles in Figures \ref{fig:brightandfaint_t1agn_2d}.
In these 2D {\sc Hi} profiles, The brightest sub-sample of All-AGN-L1 (the faintest sub-sample of All-AGN-L3) shows the weakest (the strongest) Ly$\alpha$ transmission fluctuations around the source position, $D=0$.

\begin{figure}[ht!]
\begin{center}
\includegraphics[scale=0.48]{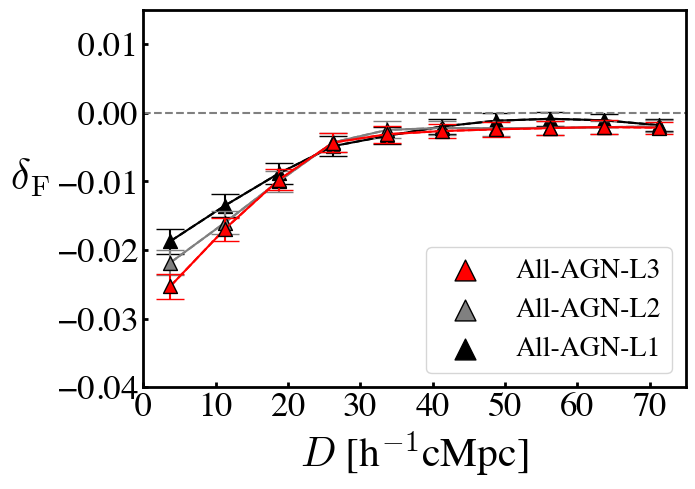}
\end{center}
\caption{Same as Figure \ref{fig:hetdext1agn_sdsst1agn}, but for the All-AGN-L3 (red), All-AGN-L2 (gray) and All-AGN-L1 (black) samples.}
\label{fig:bright_faint_t1agn_1d}
\end{figure}

We then extract the {\sc Hi} radial profiles from the 2D {\sc Hi} profiles of the All-AGN sub-samples, and present the {\sc Hi} radial profiles in Figure \ref{fig:bright_faint_t1agn_1d}.
In this figure, we find that the peak values of $\delta_{\rm F}$ for the All-AGN sub-samples is anti-correlates with AGN luminosities.
The peak $\delta_{\rm F}$ values near the source position drops from the faintest All-AGN-L3 subsample to the brightest All-AGN-L1 subsample. 
The gas densities around bright AGN are higher than (or comparable to) those around faint AGN, this result would suggest that the ionization fraction of the hydrogen gas around bright AGN is higher than the one around faint AGN on average.

\begin{figure}[ht!]
\begin{center}
\includegraphics[scale=0.48]{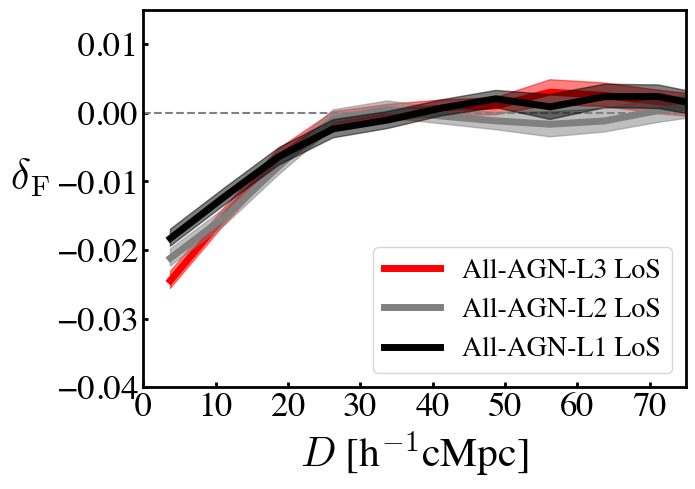}
\includegraphics[scale=0.48]{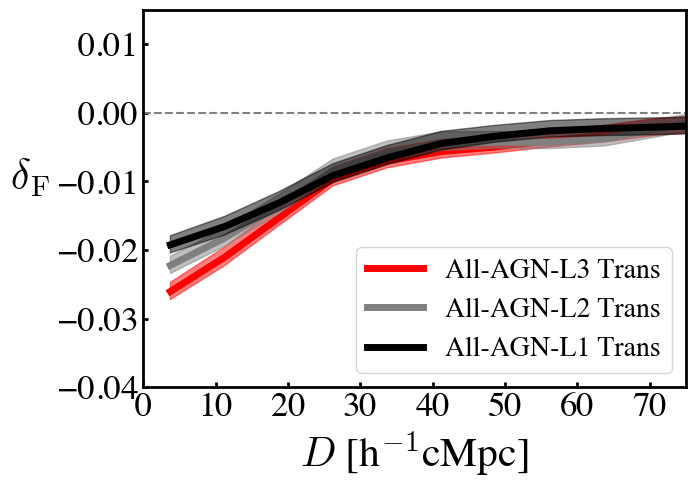}
\end{center}
\caption{LoS and Transverse {\sc Hi} radial profiles of the All-AGN-L3, All-AGN-L2, and All-AGN-L1 sub-samples. The top figure (bottom figure) presents the LoS (Transverse) {\sc Hi} radial profiles of the All-AGN-L3, All-AGN-L2, and All-AGN-L1 sub-samples, shown by the red, gray, and black lines, respectively.
The meaning of the horizontal dashed lines both in the top and bottom figures are the same as the one in Figure \ref{fig:ravoux+20}.}
\label{fig:allagn_L321_los&trans}
\end{figure}

We also present the LoS and Transverse {\sc Hi} radial profiles of the All-AGN sub-samples derived by the same method as that for the All-AGN sample in Figure \ref{fig:allagn_L321_los&trans}.
Similar to what we found in the comparison of the {\sc Hi} radial profiles for the All-AGN sub-samples, the peak values of the LoS and Transverse {\sc Hi} profiles also decrease from the faintest sub-sample, All-AGN L3, to the brightest sub-sample, All-AGN L1.
For the LoS (Transverse) {\sc Hi} radial profiles at the scales beyond 25 $h^{-1}$ cMpc, we do not find any significant differences in the comparison of the LoS (Transverse) {\sc Hi} radial profiles for the All-AGN sub-samples.

\subsubsection{AGN Type Dependence} \label{subsubsec:type_dependence}

\begin{figure}[ht!]
\begin{center}
\includegraphics[scale=0.6]{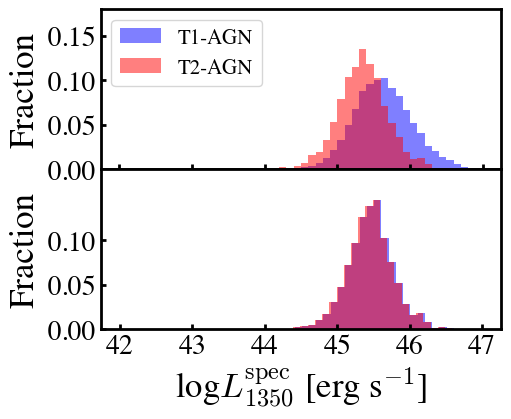}
\end{center}
\caption{Same as Figure \ref{fig:pdf_sdsst1agn_hetdext1agn}, but for the T1-AGN (blue) and T2-AGN (red) samples.}
\label{fig:t1agnt2agn_pdf}
\end{figure}

\begin{figure}[ht!]
\begin{center}
\includegraphics[scale=0.54]{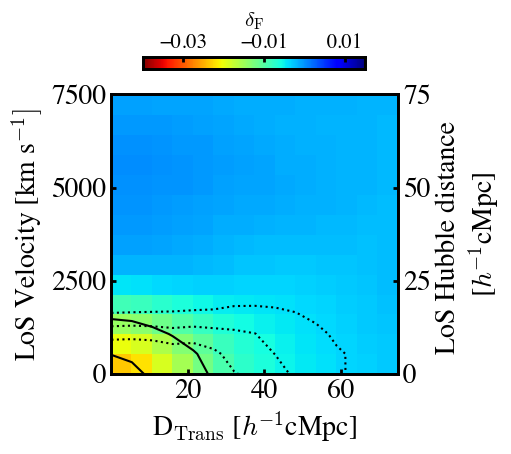}
\includegraphics[scale=0.54]{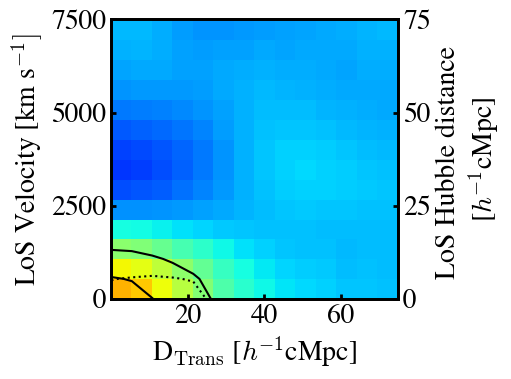}
\end{center}
\caption{Same as Figure \ref{fig:all_agn_2d}, but for the T1-AGN (top figure) and T2-AGN (bottom figure) sub-samples.}
\label{fig:t1agnt2agn_2d}
\end{figure}

\begin{figure}[ht!]
\begin{center}
\includegraphics[scale=0.48]{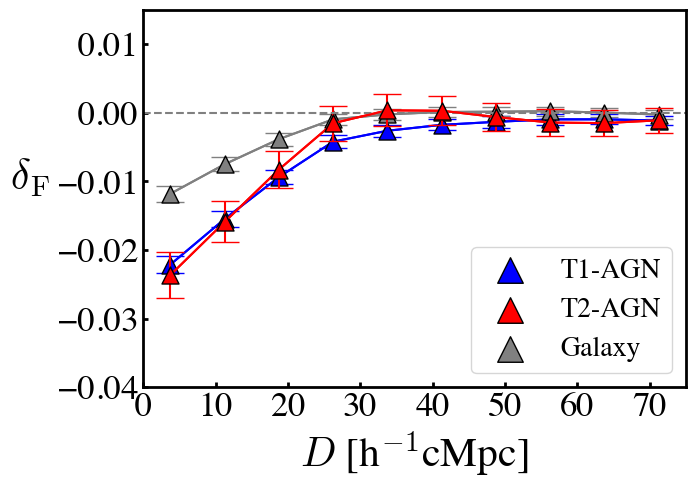}
\end{center}
\caption{Same as Figure \ref{fig:hetdext1agn_sdsst1agn}, but for the T1-AGN (blue) and T2-AGN (red) sub-samples and the galaxy (gray) sample.}
\label{fig:t1agnt2agn_1d}
\end{figure}

We investigate the dependence of {\sc Hi} profiles on type-1 and type-2 AGN.
To remove the effects of the AGN luminosity dependence (Section \ref{subsubsec:luminosity_dependence}), we make sub-samples of T1-AGN and T2-AGN with the same $L^\mathrm{spec}_\mathrm{1350}$ distribution by the same manner as the one we conduct for the selection of T1-AGN and T1-AGN(H) sub-samples in Section \ref{subsec:average_profile}.
The top panel of Figure \ref{fig:t1agnt2agn_pdf} presents the $L^\mathrm{spec}_\mathrm{1350}$ distributions of T1-AGN and T2-AGN samples, while the bottom panel of Figure \ref{fig:t1agnt2agn_pdf} shows those of the T1-AGN and T2-AGN sub-samples.
The sub-samples of T1-AGN and T2-AGN are composed of 10329 type-1 AGN and 1462 type-2 AGN, respectively.
We derive the 2D {\sc Hi} profiles from the T1-AGN and T2-AGN sub-samples. The profiles are presented in
Figure \ref{fig:t1agnt2agn_2d}.
We find $17.7$ and $7.9$ $\sigma$ detections at the source center position (0,0) of the T1-AGN and T2-AGN sub-samples, respectively.
We calculate the {\sc Hi} radial profiles from the 2D {\sc Hi} profiles of the T1-AGN and T2-AGN sub-samples. In Figure \ref{fig:t1agnt2agn_1d}, we compare the {\sc Hi} radial profiles of the T1-AGN and T2-AGN sub-samples.
No notable difference is found within 1$\sigma$ error.
The peak value of $\delta_{\rm F}=0$ of the T2-AGN subsample is within $1\sigma$ error of the peak value of the T1-AGN subsample near the source position.

\begin{figure}[ht!]
\begin{center}
\includegraphics[scale=0.48]{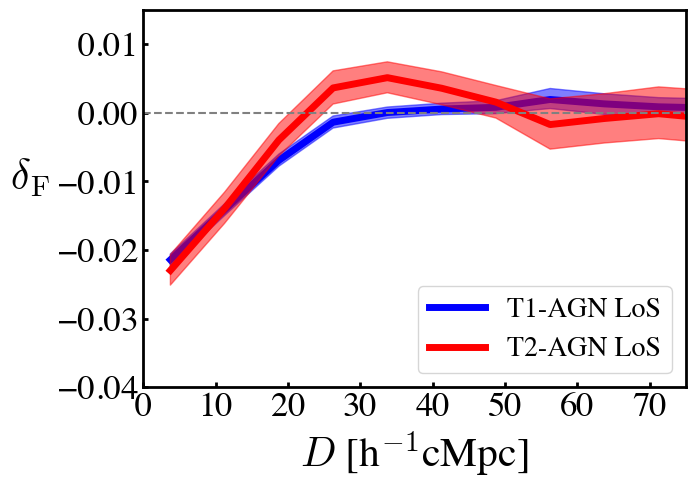}
\includegraphics[scale=0.48]{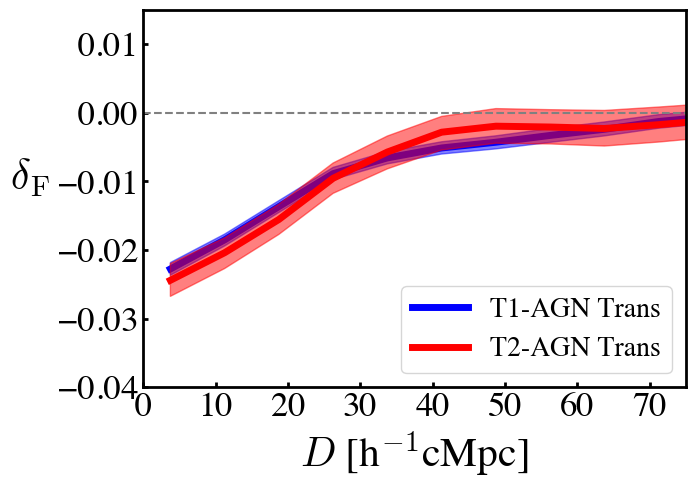}
\end{center}
\caption{Same as Figure \ref{fig:allagn_L321_los&trans}, but for the T1-AGN and T2-AGN sub-samples.}
\label{fig:t1agnt2agn_1d_LosTrans}
\end{figure}

To compare the Ly$\alpha$ forest absorptions of type-1 and type-2 AGN in the LoS and transverse directions, we derive the LoS and Transverse {\sc Hi} radial profiles of the T1-AGN and T2-AGN sub-samples and present
the profiles in Figure \ref{fig:t1agnt2agn_1d_LosTrans}.
Similar to the trend of the {\sc Hi} radial profiles, the peak values of the LoS and Transverse {\sc Hi} radial profiles for T1-AGN and T2-AGN sub-samples are not significantly different.
The comparable peak values of the LoS and Transverse {\sc Hi} radial profiles suggest that the selectively different orientation and opening angles of the dusty tori of the type-1 and type-2 AGN do not significantly affect the Ly$\alpha$ forest absorption at the scale $\lesssim15$ $h^{-1}\mathrm{cMpc}$ or our measurement dose not have enough sensitivity to detect the difference of Ly$\alpha$ forest absorption between type-1 and type-2 AGN.

For the {\sc Hi} radial profiles at the scale $>15$ $h^{-1}\mathrm{cMpc}$, we find that the $\delta_{\rm F}$ value for the LoS {\sc Hi} radial profile of the T1-AGN sub-sample is smaller than those of the T2-AGN sub-sample over the 1$\sigma$ error bar at the scale around $25$ $h^{-1}\mathrm{cMpc}$.
This result may hint that the type-2 AGN 
have a stronger power of ionization at $25$ $h^{-1}\mathrm{cMpc}$ than the type-1 AGN.
The interpretation of ionization at large-scales is in Section \ref{sec:fr+13}.

\subsection{Average HI Profiles around Galaxy} 
\label{subsec:average_profile_gy}

\begin{figure}[ht!]
\begin{center}
\includegraphics[scale=0.54]{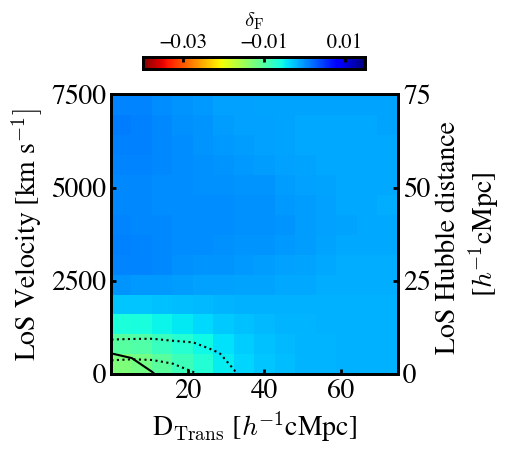}
\end{center}
\caption{2D {\sc Hi} profile of the galaxy sample sources. The color map indicates the $\delta_\mathrm{F}$ values of each cell of the 2D {\sc Hi} profile. The dotted lines show confidence level contours of $3\sigma$ and $6\sigma$. The solid line presents the contour where $\delta_\mathrm{F}$ $=-0.01$.}
\label{fig:all_hetdex_2d}
\end{figure}

We derive the 2D {\sc Hi} profile at the positions of the galaxy sample sources in the same manner as the one of the All-AGN sample sources. Figure \ref{fig:all_hetdex_2d} presents the 2D {\sc Hi} profile of the galaxy sample sources. There is a clear $10.5\sigma$ detection at the source position of (0,0).
Similarly, we calculate the {\sc Hi} radial profile from the 2D {\sc Hi} profile of the galaxy sample (Figure \ref{fig:galaxyandt1agnh}).
The {\sc Hi} radial profile of the galaxy sample shows a trend similar to those of the All-AGN sample. Both for the galaxy and All-AGN samples, the {\sc Hi} radial profile decreases towards the large scales, reaching cosmic average.

In Figure \ref{fig:all_hetdex_2d}, we find that the Ly$\alpha$ forest absorptions in the LoS and transverse directions are different.
A similar difference between the values of $A_{\rm F}$ in LoS and transverse directions of 2D {\sc Hi} profiles is claimed by \cite{mukae+20}. 
To investigate the difference between the Ly$\alpha$ forest absorptions in LoS and transverse directions for the galaxy sample, we present the LoS and Transverse {\sc Hi} radial profiles of the galaxy sample in Figure \ref{fig:galaxy_t1agnh_L&T}.
We find that the LoS and Transverse {\sc Hi} radial profiles of the galaxy sample show different gradient of the increasing $\delta_{\rm F}$ at the scale D $\sim
3.75-50$ $h^{-1}$cMpc.
This difference can be explained by the gas version of the Kaiser effect that we discussed in Section \ref{subsec:AGN_LoSandTrans_profile}.
In the LoS {\sc Hi} radial profile of the galaxy sample, we find that the $\delta_{\rm F}$ values are positive on the scale of D $=25-70$ $h^{-1}$cMpc, which is similar to the positive $\delta_{\rm F}$ values we found on the large scale of the LoS {\sc Hi} radial profile for the All-AGN sample.
We discuss these positive $\delta_{\rm F}$ values on the LoS {\sc Hi} radial profile of the galaxy sample in Section \ref{sec:fr+13}.

\subsubsection{Galaxy-AGN Dependence} \label{subsec:galaxies_agn_dependence}

\begin{figure}[ht!]
\begin{center}
\includegraphics[scale=0.54]{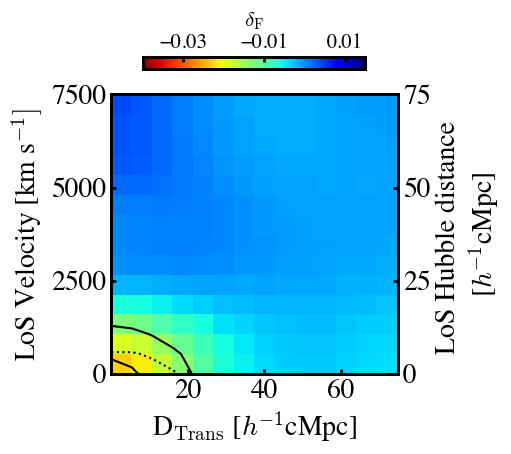}
\end{center}
\caption{Same as Figure \ref{fig:brightandfaint_t1agn_2d}, but for the T1-AGN(H) sample. }
\label{fig:galaxyandt1agnh}
\end{figure}

\begin{figure}[ht!]
\begin{center}
\includegraphics[scale=0.48]{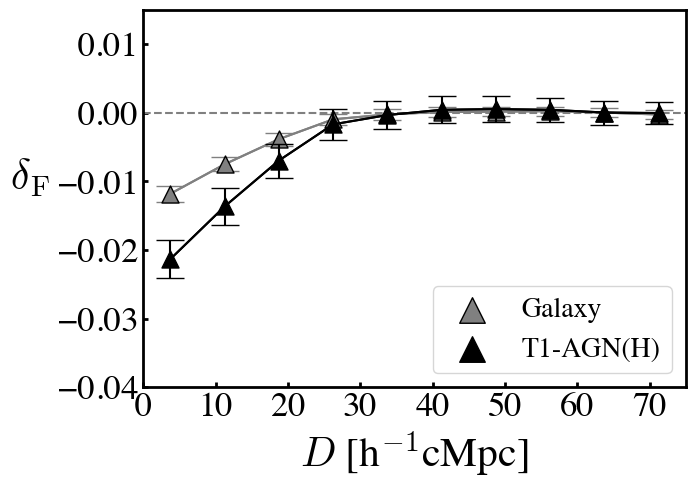}
\end{center}
\caption{Same as Figure \ref{fig:bright_faint_t1agn_1d}, but for galaxy (gray) and T1-AGN(H) (black) samples.}
\label{fig:galaxy_t1agnh_1dprofile}
\end{figure}

We derive 2D {\sc Hi} profiles for the T1-AGN(H) sample constructed from the HETDEX data.
Figure \ref{fig:all_hetdex_2d} and \ref{fig:galaxyandt1agnh} show the 2D {\sc Hi} profiles of the galaxy and T1-AGN(H) samples. 
We find $7.6$$\sigma$ detection around the source position for the T1-AGN(H) sample.
Figure \ref{fig:galaxy_t1agnh_1dprofile} presents the {\sc Hi} radial profiles of the galaxy and T1-AGN(H) samples derived from the 2D {\sc Hi} profiles. We also compare the {\sc Hi} radial profiles of the galaxy sample with those of T1-AGN and T2-AGN in Figure \ref{fig:t1agnt2agn_1d}.
In the {\sc Hi} radial profiles of the galaxy and T1-AGN(H) samples, the $\delta_{\rm F}$ values decrease toward the source position $D=0$.
In Figure \ref{fig:galaxy_t1agnh_1dprofile} (\ref{fig:t1agnt2agn_1d}), we find that the $\delta_{\rm F}$ values of T1-AGN(H) (T1-AGN and T2-AGN) are smaller than those of the galaxies at $\lesssim 20$ $h^{-1}$ cMpc. These $\delta_{\rm F}$ excesses of the AGN may be explained by the hosting dark matter halos of the AGN being more massive than those of the galaxies.
\citet{momose+21} also investigate the {\sc Hi} radial profile around AGN, and find Ly$\alpha$ forest absorption decrement at the source center ($\lesssim5$ $h^{-1}$Mpc). They argure that this trend can be explained by the proximity effect. On the other hand, their result is different from ours that the $A_{\rm F}$ values monotonically increase with decreasing distance.
This difference between our and \citeauthor{momose+21}'s results is produced by the fact that our results for $\lesssim 10$ $h^{-1}$ cMpc are largely affected by the Ly$\alpha$ transmission fluctuation at $\sim 10$ $h^{-1}$ cMpc due to the coarse resolution of our {\sc Hi} tomography map, 15 $h^{-1}$ cMpc, in contrast with $2.5$ $h^{-1}$ cMpc for the resolution of \cite{momose+21}.

\begin{figure}[ht!]
\begin{center}
\includegraphics[scale=0.48]{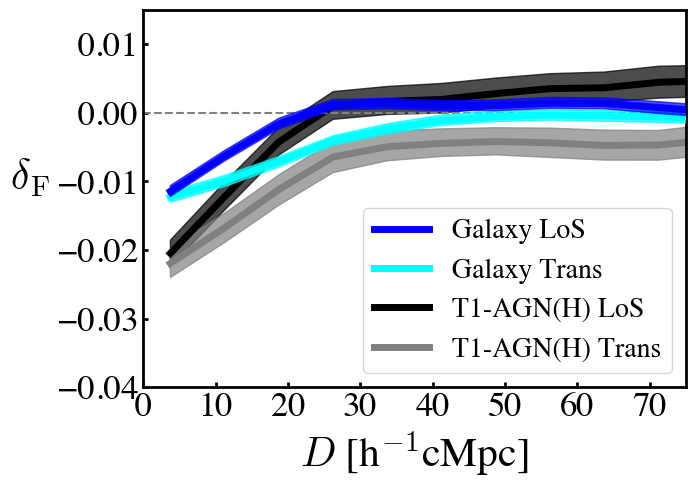}
\end{center}
\caption{Same as Figure \ref{fig:all_agn_LT}, but for the galaxy and T1-AGN(H) samples.}
\label{fig:galaxy_t1agnh_L&T}
\end{figure}

We then derive the LoS and Transverse radial {\sc Hi} profile of the T1-AGN(H) sample.
The results of the profiles are shown in Figure \ref{fig:galaxy_t1agnh_L&T}.
Similar to the LoS and Transverse {\sc Hi} radial profiles of the All-AGN and galaxy samples, the gas version of the Kaiser effect and the positive $\delta_{\rm F}$ in the LoS direction on the scale beyond $D=25$ h$^{-1}$cMpc are also found in those of the T1-AGN(H) sample.

\subsection{Comparison with Theoretical Models}\label{sec:fr+13}

There are theoretical models of {\sc Hi} radial profiles around AGN that are made by \citet{FR+13}. \citet{FR+13} present their {\sc Hi} radial profiles with 
the LoS distance
in the form of cross-correlation function (CCF).

We first calculate theoretical CCFs of All-AGN, following the definition of the CCF presented in \citet{FR+12,FR+13}.
\citet{FR+13} assume the linear cross-power spectrum of the QSOs and Ly$\alpha$ forest,
\begin{equation} \label{eq:ccf0}
    P_{\rm qF}(\mathbf{k},z)=b_{\rm q}(z)[1+\beta_{\rm q}(z) \mu^2_{\rm k}]b_{\rm F}(z)[1+\beta_{\rm F}(z)\mu^2_{\rm k}]P_{\rm L}(k,z),
\end{equation}
where $P_{\rm L}(k,z)$ is the linear matter power spectrum.
Here $\mu_{\rm k}$ is the cosine of the angle between the Fourier mode and the LoS \citep{Kaiser+87}. 
The values of $b_{\rm q}$ and $b_{\rm F}$ ($\beta_{\rm q}$ and $\beta_{\rm F}$) are the bias factors (redshift space distortion parameters) of the QSO and Ly$\alpha$ density, respectively.

The redshift distortion parameter of QSO obeys the relation $\beta_{\rm q}=f(\Omega)/b_{\rm q}$, where $f(\Omega)$ is the logarithmic derivative of the linear growth factor \citep{Kaiser+87}, $b_{\rm q}=3.8\pm0.3$ \citep{White+12}.
We use the condition of Ly$\alpha$ forest, $b_{\rm F}(1+\beta_{\rm F})=-0.336$ for $b_F \propto (1+z)^{2.9}$,
that is determined by observations of Ly$\alpha$ forest at $z\simeq 2.25$ \citep{Slosar+11}.
\citet{FR+13} estimate the CCF of QSOs by the Fourier transform of $P_{\rm qF}$ \citep{Hamilton+92}:
\begin{equation} \label{eq:ccf1}
    \xi({\bf r})=\xi_0(r)P_0(\mu)+\xi_2(r)P_2(\mu)+\xi_4(r)P_4(\mu),
\end{equation}
where $\mu$ is the cosine of angle between the position ${\bf r}$ and the LoS in the redshift space. The values of $P_0$, $P_2$, and $P_4$ are the Legendre polynomials, $P_0=1$, $P_2=(3\mu^2-1)$, and $P_4=(35\mu^4-30\mu^2+3)/8$, respectively.
The functions of $\xi_0$, $\xi_2$, and $\xi_4$ are:
\begin{equation} \label{eq:ccf2}
    \xi_0(r)=b_{\rm q} b_{\rm F} [1+(\beta_{\rm q}+\beta_{\rm F})/3+\beta_{\rm q}\beta_{\rm F}/5]\zeta(r),
\end{equation}
\begin{equation} \label{eq:ccf3}
    \xi_2(r)=b_{\rm q} b_{\rm F} [2/3(\beta_{\rm q}+\beta_{\rm F})+4/7\beta_{\rm q}\beta_{\rm F}][\zeta(r)-\bar{\zeta}(r)],
\end{equation}
\begin{equation} \label{eq:ccf4}
    \xi_4(r)=8/35 b_{\rm q} b_{\rm F} \beta_{\rm q}\beta_{\rm F} [\zeta(r)-5/2\bar{\zeta}(r)-7/2\bar{\bar{\zeta}}(r)].
\end{equation}
The function $\zeta(r)$ is the standard CDM linear correlation function in real space \citep{Bardeen+86,Hamilton+91}.
The functions $\bar{\zeta}(r)$ and $\bar{\bar{\zeta}}(r)$ are given by:
\begin{equation} \label{eq:ccf5}
    \bar{\zeta}(r) \equiv 3r^{-3} \int^r_0 \zeta(s)s^2ds ,
\end{equation}
\begin{equation} \label{eq:ccf6}
    \bar{\bar{\zeta}}(r) \equiv 5r^{-5} \int^r_0 \zeta(s)s^4ds.
\end{equation}
In Figure \ref{fig:allagn_model},
we present $D\xi$ as a function of the LoS distance for the model of \citet{FR+13} that is calculated under the assumption of the mean overdensity of the $15$ $h^{-1}$cMpc corresponding to the spatial resolution of our observational results.

To compare our observational measurements with the model CCF of \citet{FR+13}, we calculate the value of $\xi$ for our All-AGN sample.
The value of $\xi$ in each cell $\xi_{\rm cell}$ 
is calculated by
\begin{equation} \label{eq:cross-correlation}
    \xi_{\rm cell} = \frac{\sum_{i\in {\rm cell}} \omega_{i} \delta_{{\rm F}i}}{\sum_{i\in {\rm cell}} \omega_{i}},
\end{equation}
where $\omega_{i}$ is the weight determined by the observational errors and the intrinsic variance of the Ly$\alpha$ forest. Noted that the $\delta_{\rm Fi}$ used in \cite{FR+12,FR+13} is the raw $\delta_{\rm Fi}$, which is not undergoing the Wiener filtering scheme. The value of $\omega_{i}$ is obtained by
\begin{equation} \label{eq:weight}
    \omega_{i} = \left\lbrack \sigma^2_{\rm F}(z_i)+\frac{1}{\langle S/N \rangle^2\times \langle F(z_i) \rangle ^2 } \right\rbrack^{-1},
\end{equation}
where $\sigma_{\rm F}(z_i)$ is the intrinsic variance of the Ly$\alpha$ forest. The value of $\langle F(z_i) \rangle $ is the cosmic average Ly$\alpha$ transmission (Eq.\ref{eq:mf}).
We adopt $\langle S/N \rangle = 1.4$ that is the criterion of the background source selection (Section \ref{subsec:bkagn}).
The intrinsic variance, $\sigma_{\rm F}(z_i)$, of the Ly$\alpha$ forest taken from \cite{FR+13} is:
\begin{equation} \label{eq:intrinsic variance}
    \sigma^2_{\rm F}(z_i) = 0.065[(1+z_i)/3.25]^{3.8}.
\end{equation}

We calculate $\xi$ with our All-AGN sample via
the Equations \ref{eq:cross-correlation}, \ref{eq:weight}, and \ref{eq:intrinsic variance}, using
the binning sizes same as those in \cite{FR+13}.
We present $\xi$ multiplied by $D$ with the black squares in Figure \ref{fig:allagn_model}.
(explanation of Momose+21)
For reference, we also derive the $\xi$ for our galaxy sample shown by the blue triangles.

In Figure \ref{fig:allagn_model}, we find that the $D\xi$ profile of our All-AGN sample show a trend similar to the one of the model predicted by \citet{FR+13}.
The observational $D\xi$ profile of our All-AGN sample shows a good agreement with the model $D\xi$ profile of \cite{FR+13} at the scale of $D>30$ $h^{-1}$cMpc.
Equation \ref{eq:ccf0} of the linear theory model already includes the parameter of redshift distortion, $\beta$, which is due to the coherent motions of the {\sc Hi} gas around the quasars. 
The model is a prediction on the impact of the clustering effect that a quasar statistically gathers {\sc Hi} gas from large scales, even $\gtrsim 30$ $h^{-1}$cMpc. The positive $D\xi$ structure $\gtrsim 30$ $h^{-1}$cMpc can be explained by `cosmic voids' like structure whose {\sc Hi} gas column density is slightly smaller than the cosmic average.
Though the general trend of the positive $D\xi$ structure of our results at $\gtrsim 30$ $h^{-1}$cMpc are the same as the model $D\xi$ profile, the model $D\xi$ profile is slightly lower than the $D\xi$ profiles of the observations at $\gtrsim 60$ $h^{-1}$cMpc. We can not rule out the possibility that the ionization of AGN make an extra decreasing on the Ly$\alpha$ absorption at $\gtrsim 60$ $h^{-1}$cMpc.
\cite{FR+13} also present the model of ionization. In the model of ionization, \cite{FR+13} assume the spectrum of the AGN at $D = 0$ with $L_\nu \propto \nu^{- \alpha}$, where $\alpha = 1.5$ (1.0) for the frequency $\nu$ over (below) the Lyman limit. The luminosity of $\lambda=1420$ \AA\ is normalized as $L_\nu= 3.1\times10^{30}$ erg/s/Hz, which is taken from the mean luminosity of the SDSS data release 9 quasars.
No assumptions of AGN type have been made in the models of Font-Ribera+13.
Based on the model of ionization, \cite{FR+13} calculate $\xi$ for the homogeneous gas radiated by AGN, and obtain the function
\begin{equation} \label{eq:ionization}
    \xi= 0.0065(20\; h^{-1}{\rm cMpc}/D)^2.
\end{equation}
With the $\xi$ function, we calculate $D\xi$ that is presented with the cyan dashed curve in Figure \ref{fig:allagn_model}. The cyan dashed curve shows the plateau at $D\geq40$ $h^{-1}{\rm cMpc}$ with positive $D\xi$ values.
To distinguish the large-scale positive $D\xi$ values, which are referred to as the `weak absorption outskirts', from the proximity zone created by the proximity effect, we plot the observational CCF of AGN obtained by \cite{momose+21} in Figure \ref{fig:allagn_model}. The AGN CCF obtained by \citeauthor{momose+21} shows a decreasing Ly$\alpha$ forest absorption toward source position ($D=0$ $h^{-1}$cMpc) caused by the proximity effect. 
If the weak absorption outskirts are created by the combination of the clustering effect and ionization, 
our findings indicate that the {\sc Hi} radial profile of AGN may has transitions from proximity zones ($\lesssim$ a few $h^{-1}$cMpc) to the {\sc Hi} structures ($\sim 1-30$ $h^{-1}$cMpc) and the ionized outskirts ($\gtrsim 30$
$h^{-1}$cMpc). The hard radiation may pass through the {\sc Hi} structure due to the small cross-section and ionizes the {\sc Hi} gas in the regions of ionized outskirts. Because of the low recombination rate, the {\sc Hi} gas remains ionized in the weak absorption outskirts.

Figure \ref{fig:allagn_model} shows that the $D\xi$ values in the range of {\sc Hi} structure around AGN and galaxy are also similar.
Interestingly, the $D\xi$ profile of our galaxy sample also shows positive $D\xi$ values towards $\gtrsim 30$ $h^{-1}$cMpc which is similar to those of the AGN model and our All-AGN sample.
This result may suggest that the {\sc Hi} gas at large scale ($\gtrsim 20$ $h^{-1}$cMpc) around galaxies also fall toward the source position ($D=0$).
Regions around galaxies are special as galaxies are clustered together.
Galaxies in this work are bright with $M_{\rm UV}<-22$ mag. The galaxies can be hosted by massive haloes, and are likely to distribute at overdensity regions.
The overdensity region suggests that each galaxy can be surrounded by several galaxies.
Although it is difficult for a galaxy to trigger the clustering effect for the {\sc Hi} gas on a scale of $\gtrsim 20$ $h^{-1}$cMpc, a group of galaxies may have enough gravitational power to aggregate the {\sc Hi} on this scale.

\begin{figure}[ht!]
\begin{center}
\includegraphics[scale=0.46]{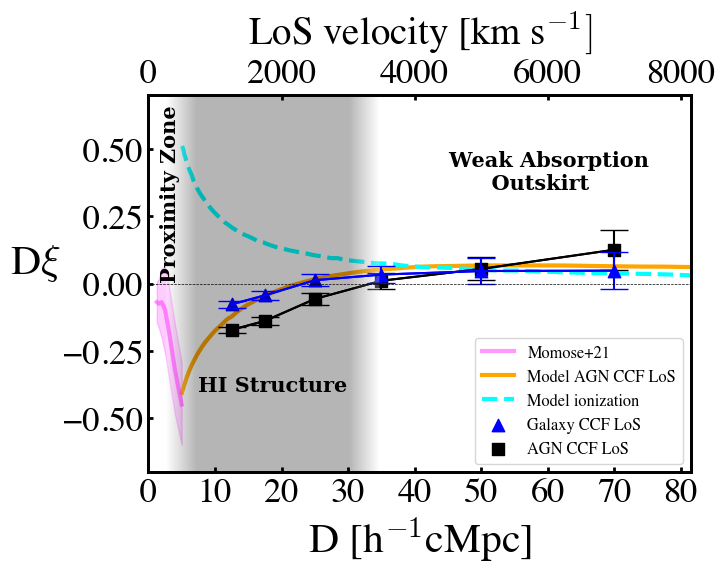}
\end{center}
\caption{Comparison between our All-AGN and galaxy results and the models of \cite{FR+13} in the LoS CCF $\xi$ multiplied by distance ($D$). The black and blue points are the results derived from the All-AGN and galaxy samples sources, respectively. The orange curve is the LoS CCF of QSOs with the Ly$\alpha$ forest derived by \cite{FR+13}. The cyan dashed curve shows the ionization of radiation effect taken from \cite{FR+13}. The pink line presents the CCF of AGN obtained by \cite{momose+21}. The gray shade presents the range of the {\sc Hi} structure. Two white areas show the regions of proximity zone and weak absorption outskirts. The horizontal gray line indicates the cosmic average where $D\xi=0$.}
\label{fig:allagn_model}
\end{figure}

\section{Summary}

We reconstruct two 3D {\sc Hi} tomography maps based on the Ly$\alpha$ forests in the spectra of 14763 background QSOs from the SDSS survey with no signatures of damped Ly$\alpha$ system or broad absorption lines.
The maps cover the extended Fall and Spring fields defined by the HETDEX survey.
The spatial volume of the reconstructed 3D {\sc Hi} tomography maps are $2257 \times 233 \times 811$ $h^{\rm -3}$cMpc$^3$ and $3475 \times 1058 \times 811$ $h^{\rm -3}$cMpc$^3$.
We investigate Ly$\alpha$ forest absorption around galaxies and AGN with samples made from HETDEX and SDSS survey results in our study field.
Our results are summarized below.
\begin{itemize}
    \item We derive the 2D {\sc Hi} and {\sc Hi} radial profiles of the All-AGN sample consisted of SDSS AGN. We find that the 2D {\sc Hi} profile is more extended in the transverse direction than along the line of sight. In the {\sc Hi} radial profile All-AGN sample, the values of Ly$\alpha$ transmission fluctuation, $\delta_{\rm F}$, increase toward the large scale, touching to $\delta_{\rm F} \sim 0$.
    
    \item We compare the {\sc Hi} radial profiles derived from the T1-AGN and T1-AGN(H) sub-samples, whose $L^{\rm spec}_{\rm 1350}$ distributions are the same.
    We find that the {\sc Hi} radial profile of the T1-AGN sub-sample agrees with that of the T1-AGN(H) sub-sample. 
    This agreement suggests that the systematic uncertainty between the SDSS and the HETDEX survey results is negligible.
    
    \item We examine the dependence of the {\sc Hi} profile on AGN luminosity by deriving the 2D {\sc Hi}, {\sc Hi} radial, LoS {\sc Hi} radial, and Transverse {\sc Hi} radial profiles of the All-AGN-L3 (the faintest), All-AGN-L2, and All-AGN-L1 (the brightest) sub-samples.
    We find that the Ly$\alpha$ forest absorption is the greatest in the lowest-luminosity AGN sub-sample, and that the Ly$\alpha$ forest absorption becomes weaker with increasing AGN luminosity
    This result suggests that, on average, if the density of {\sc Hi} gas around the bright AGN is greater than (or comparable to) those of the faint AGN, the ionization fraction of {\sc Hi} gas around bright AGN is higher than that around faint AGN.
    
    \item We investigate the AGN type dependence of Ly$\alpha$ forest absorption around type-1 and type-2 AGN by the 2D {\sc Hi}, {\sc Hi} radial, LoS {\sc Hi} radial, and Transverse {\sc Hi} radial profiles extracted from the T1-AGN and T2-AGN sub-samples with the same $L^{\rm spec}_{\rm 1350}$ distributions.
    The comparison between the {\sc Hi} radial profiles of T1-AGN and T2-AGN sub-samples indicates that the Ly$\alpha$ transmission fluctuation around the T2-AGN sub-sample is comparable to the one of the T1-AGN sub-sample on average.
    This trend suggests that, the selectively different opening angle and orientation of the dusty torus for type-1 and type-2 AGN do not have a significant impact on the Mpc-scale Ly$\alpha$ forest absorption or the sensitivity of our result is not enough to detect the difference.

    \item We compare the Ly$\alpha$ forest absorptions around galaxies and type-1 AGN with the 2D {\sc Hi}, {\sc Hi} radial, LoS {\sc Hi} radial, and Transverse {\sc Hi} radial profiles derived from the galaxy and T1-AGN(H) sample sources.
    The Ly$\alpha$ transmission fluctuation values, $\delta_{\rm F}$, around the T1-AGN(H) sample are larger than those of the Galaxy sample on average.
    This result may be caused by the dark matter halos of type-1 AGN having a larger mass than the one of galaxies on average.
    
    \item We find that the {\sc Hi} radial profiles of the LoS distance for the galaxy and All-AGN samples show positive $\delta_{\rm F}$ values, which means weak Ly$\alpha$ forest absorption, at the scale over $\sim 30$ $h^{-1}$cMpc. We extract the $D\xi$ profile of our galaxy and All-AGN samples to compare with the model CCF of AGN from \cite{FR+13}. The general trend of the positive $D\xi$ at $\gtrsim 30$ $h^{-1}$cMpc is the same as the model CCF.
    This results suggest that the {\sc Hi} radial profile of AGN has transitions from proximity zones ($\lesssim$ a few $h^{-1}$cMpc) to the {\sc Hi} rich structures ($\sim 1-30$ $h^{-1}$cMpc) and the weak absorption outskirts ($\gtrsim 30$ $h^{-1}$cMpc).

\end{itemize}

\section*{Acknowledgements}
We thank Nobunari Kashikawa, Khee-Gan Lee, Akio Inoue, Rikako Ishimoto, Shengli Tang, Yongming Liang, Rieko Momose, and Koki Kakiichi for giving us helpful comments.

HETDEX is led by the University of Texas at Austin McDonald Observatory and Department of Astronomy with participation from the Ludwig-Maximilians-Universität München, Max-Planck-Institut für Extraterrestrische Physik (MPE), Leibniz-Institut für Astrophysik Potsdam (AIP), Texas A\&M University, Pennsylvania State University, Institut für Astrophysik Göttingen, The University of Oxford, Max-Planck-Institut für Astrophysik (MPA), The University of Tokyo and Missouri University of Science and Technology. In addition to Institutional support, HETDEX is funded by the National Science Foundation (grant AST-0926815), the State of Texas, the US Air Force (AFRL FA9451-04-2- 0355), and generous support from private individuals and foundations.
The observations were obtained with the Hobby-Eberly Telescope (HET), which is a joint project of the University of Texas at Austin, the Pennsylvania State University, Ludwig-Maximilians-Universität München, and Georg-August-Universität Göttingen. The HET is named in honor of its principal benefactors, William P. Hobby and Robert E. Eberly.
The authors acknowledge the Texas Advanced Computing Center (TACC) at The University of Texas at Austin for providing high performance computing, visualization, and storage resources that have contributed to the research results reported within this paper. URL: http://www.tacc.utexas.edu

VIRUS is a joint project of the University of Texas at Austin,
Leibniz-Institut f\"ur Astrophysik Potsdam (AIP), Texas A\&M University
(TAMU), Max-Planck-Institut f\"ur Extraterrestrische Physik (MPE),
Ludwig-Maximilians-Universit\"at Muenchen, Pennsylvania State
University, Institut fur Astrophysik G\"ottingen, University of Oxford,
and the Max-Planck-Institut f\"ur Astrophysik (MPA). In addition to
Institutional support, VIRUS was partially funded by the National
Science Foundation, the State of Texas, and generous support from
private individuals and foundations.

This work is supported in part by MEXT/JSPS KAKENHI Grant Number 21H04489 (HY), JST FOREST Program, Grant Number JP-MJFR202Z (HY).

K. M. acknowledges financial support from the Japan Society for the Promotion of Science (JSPS) through KAKENHI grant No. 20K14516.

This paper is supported by World Premier International
Research Center Initiative (WPI Initiative), MEXT, Japan, the
joint research program of the Institute of Cosmic Ray Research (ICRR), the University of Tokyo, and KAKENHI (19H00697,
20H00180, and 21H04467) Grant-in-Aid for Scientific
Research (A) through the Japan Society for the Promotion of
Science.


\clearpage
\appendix


\begin{figure*}[ht!]
\begin{center}
\includegraphics[scale=0.55]{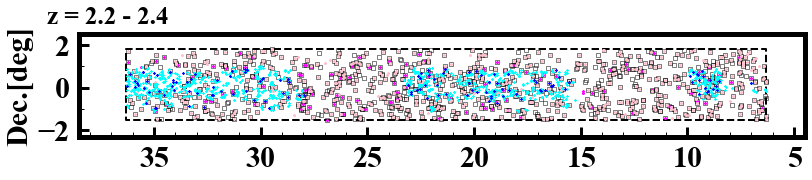}
\includegraphics[scale=0.55]{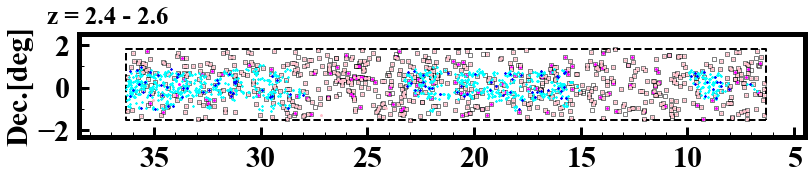}
\includegraphics[scale=0.55]{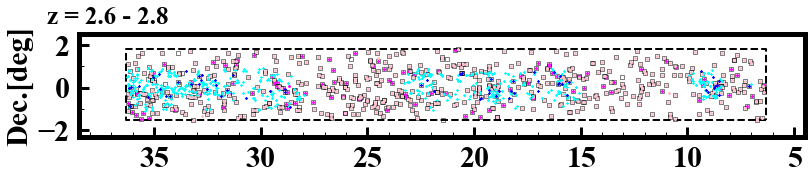}
\includegraphics[scale=0.55]{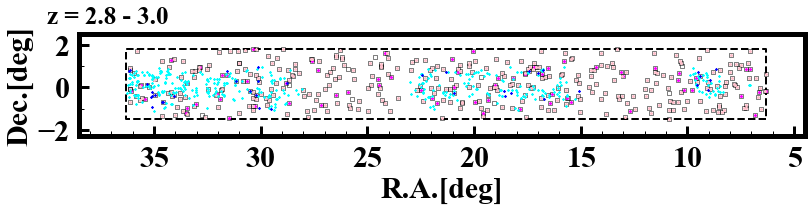}
\end{center}
\caption{Continued from Figure \ref{fig:our_study_field}. The different panels denote the coverages over different redshift ranges shown at the top left of each panel.}
\label{fig:our_study_field_ap}
\end{figure*}

\begin{figure*}[ht!]
\begin{center}
\includegraphics[scale=0.55]{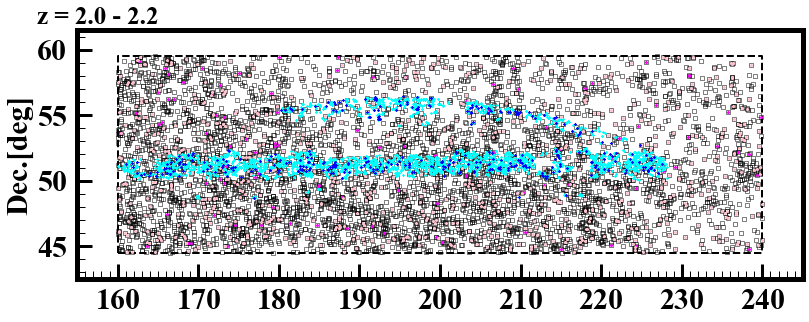}
\includegraphics[scale=0.55]{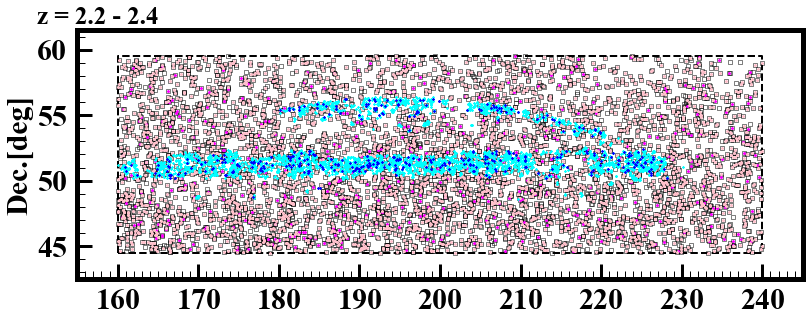}
\includegraphics[scale=0.55]{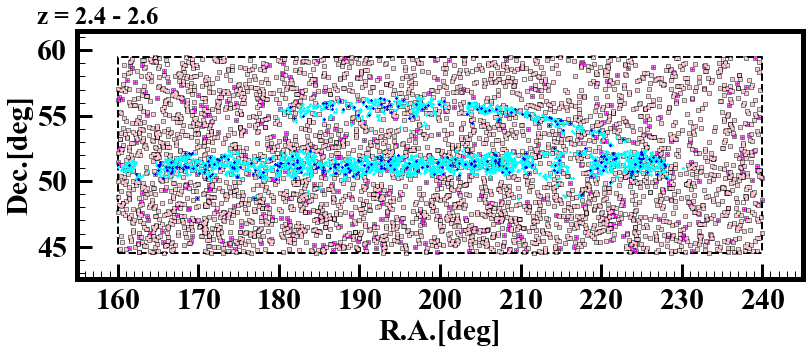}
\end{center}
\caption{
Same as Figure \ref{fig:our_study_field}, but for the foreground sources in the ExSpring field.}
\label{fig:exspring_fg_1}
\end{figure*}

\begin{figure*}[ht!]
\begin{center}
\includegraphics[scale=0.55]{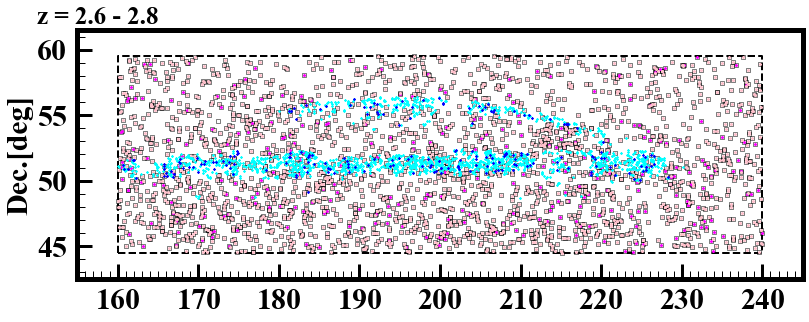}
\includegraphics[scale=0.55]{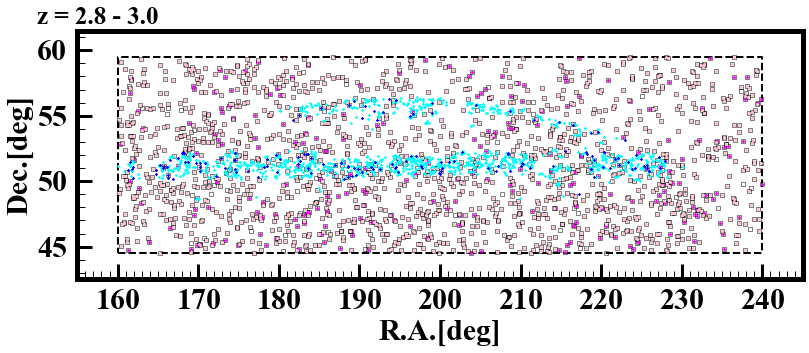}
\end{center}
\caption{Continued from Figure \ref{fig:exspring_fg_1}.}
\label{fig:exspring_fg_2}
\end{figure*}

\begin{figure*}[ht!]
\begin{center}
\includegraphics[scale=0.55]{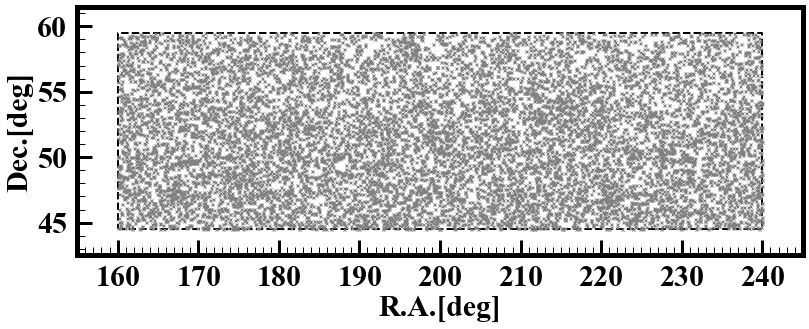}
\end{center}
\caption{
Same as Figure \ref{fig:exfall_bk}, but for the background sources in the ExSpring field.
}
\label{fig:exspring_bk}
\end{figure*}

\end{document}